\documentclass[comsoc, journal]{IEEEtran}
\IEEEoverridecommandlockouts
\usepackage[numbers]{natbib}
\setcitestyle{open={[},close={]}}

\usepackage[cmintegrals]{newtxmath}
\usepackage{amsmath}%,amssymb}%,amsfonts}

\usepackage{graphicx}
\usepackage{textcomp}
\usepackage{xcolor}
\usepackage{float}
\usepackage{array}
\usepackage[font=footnotesize,labelfont=bf]{caption}
\usepackage{optidef}
\usepackage{subcaption}
\usepackage[utf8]{inputenc}
\usepackage{algorithm}
\usepackage{algpseudocode}
\usepackage{color, xcolor, colortbl}
\newcommand{\revision}{\color{black}}

\def\BibTeX{{\rm B\kern-.05em{\sc i\kern-.025em b}\kern-.08em
    T\kern-.1667em\lower.7ex\hbox{E}\kern-.125emX}}
\begin{document}

\title{Data-Aware Device Scheduling for Federated Edge Learning\\}

\author{\IEEEauthorblockN{
		Afaf Taïk \IEEEmembership{Student Member, IEEE}, Zoubeir Mlika \IEEEmembership{Member, IEEE}  and
		Soumaya Cherkaoui \IEEEmembership{Senior Member, IEEE} }	\\
	\IEEEauthorblockA{
		INTERLAB, Engineering Faculty, Université de Sherbrooke, Canada.\\ 
		\{afaf.taik, zoubeir.mlika, soumaya.cherkaoui\}@usherbrooke.ca }}
\maketitle

\begin{abstract}
Federated Edge Learning (FEEL) involves the collaborative training of machine learning models among edge devices, with the orchestration of a  server in a wireless edge network. Due to frequent model updates, FEEL needs to be adapted to the limited communication bandwidth, scarce energy of edge devices, and the statistical heterogeneity of edge devices' data distributions. Therefore, a careful scheduling of a subset of devices for training and uploading models is necessary. In contrast to previous work in FEEL where the data aspects are under-explored, we consider data properties at the heart of the proposed scheduling algorithm.
To this end, we propose a new scheduling scheme for non-independent and-identically-distributed (non-IID) and unbalanced datasets in FEEL. As the data is the key component of the learning,  we propose a new set of considerations for data characteristics in wireless scheduling algorithms in FEEL. In fact, the data  collected by the devices depends on the local environment and usage pattern. Thus, the datasets vary in size and distributions among the devices. In the proposed algorithm, we consider both data and resource perspectives. In addition to minimizing the completion time of FEEL as well as the transmission energy of the participating devices, the algorithm prioritizes devices with rich and diverse datasets.  We first define a general framework for the data-aware scheduling and the main axes and requirements for diversity evaluation. Then, we discuss diversity aspects and some exploitable techniques and metrics. Next, we formulate the problem and present our {\revision data-aware scheduling (DAS) algorithm} for FEEL. Evaluations in different scenarios show that DAS algorithm can help achieve high accuracy in few rounds with a reduced cost.
\end{abstract}

\IEEEpeerreviewmaketitle


\section{Introduction}
\label{sec:introduction}
Machine learning (ML) models require large and rich sets of data for training. Nonetheless, the collection of large volumes of data generated by connected devices over wireless networks raises concerns related to data privacy and network congestion {\revision \cite{reviewer1_1,reviewer1_2}}. Federated edge learning (FEEL) \cite{f3, zhu_broadband_2020} was proposed to tackle these concerns, by implementing distributed ML at the  edge of the network. In addition to preserving  privacy  by keeping the data locally, FEEL benefits from rapid access to the data generated by end devices and leveraging their computational resources. In FEEL, the model training is performed on edge devices with the orchestration of a multi-access edge computing (MEC) server. Each device trains the model using its local data, and only the resultant model parameters or stochastic gradients are sent to the MEC server for aggregation. 

The scarce resources, especially the communication bandwidth, limit the efficacy of FEEL operations, particularly for the transmission of large size models. Consequently, most of the existing works in FEEL focus on designing scheduling algorithms with optimal resource usage. Several proposed works aim, for instance, to minimize the completion time of FEEL
\cite{fedcs}, local computation energy \cite{i1}, or transmission energy of participating devices \cite{zeng_energy-efficient_2019}. As a result, the number of scheduled devices is often restricted as a means to meet latency and energy constraints. This restriction often slows down the convergence of training \cite{yang_federated_2020,zeng_energy-efficient_2019}. 
Therefore, scheduling algorithms aim to maximize the number of  collected updates in each round, but this scheduling goal can be biased towards powerful devices with smaller datasets. Thus, the collected updates might not be representative, as they are not trained on richer data. To avoid this issue, scheduling algorithms should also aim to diversify the participating devices through the use of fairness measures \cite{yang_age-based_2020, yang_scheduling_2019}. But, the amount of connected devices grows faster than the network capacities, which will make scaling these algorithms harder in practice. Moreover, Internet of things data is highly redundant and inherently unbalanced, given that the data collected by the devices depend on the local environment and the device's usage pattern.  Therefore, the size and the statistical properties of  local datasets distributions vary among devices \cite{caldas_leaf_2019}. Thus, a careful selection of participating devices imposes the consideration of their data properties, which motivates this work. 

The main idea we advocate in this paper finds its roots in active learning \cite{shi_diversifying_2016, you_diverse_2014}, where models are trained using fewer data points provided that the chosen samples are required to be diverse and informative. While in active learning, the selection concerns single unlabelled data points, the selection in FEEL concerns complete datasets with  already labelled data points, and therefore requires a different evaluation of diversity. Additionally, the incorporation of the diversity measures in FEEL requires different considerations in regards of privacy and the properties of the FEEL setting \cite{taik_federated_2020}.
In this paper, we consider diversity as the baseline criterion for choosing participating devices in FEEL. The diversity evaluation is applied on datasets, where the priority is given to devices with potentially more informative datasets to speed up the training process. To this end, we propose a method for incorporating datasets' diversity properties in FEEL scheduling, by identifying a set of dataset diversity measures, and designing a data-aware scheduling (DAS) algorithm. 

The contributions of this paper can be summarized as follows:
\begin{itemize}
    \item[1)] we design a suitable diversity indicator, which serves as a priority criterion for the selection of devices;
    \item[2)] we formulate a joint device selection and bandwidth allocation problem taking into account data diversity; 
    \item[3)]we prove that the formulated problem is NP-hard and we propose a data-aware scheduling algorithm based on an iterative decomposition technique to solve it; and
    \item[4)] we evaluate the proposed diversity indicator and the DAS algorithm through extensive simulations. 
\end{itemize}

The remainder of this paper is organized as follows. In Section II, we present the background for FEEL and related work. In Section III, we present the design of the proposed diversity measure, starting with the used uncertainty measures and their integration in FEEL. In Section IV, we integrate the proposed measure in the design of the joint selection and bandwidth allocation algorithm. Simulation results are presented in Section V. At last, conclusions and final remarks are presented in Section VI.

\section{Background and related work}
\label{sec:Related work}
In this section, we start by briefly introducing the main concepts of FEEL. Then, we describe the existing challenges in deploying FL in wireless edge networks. Next, we discuss the related work, illustrate the existing research gaps and motivate the need for a new scheduling scheme for FEEL.
\subsection{Federated Edge Learning}

In contrast to centralized training, FL keeps the training data at each device and learns a shared global model through the federation of distributed connected devices. Keeping data locally yields many benefits, reduced bandwidth use, and rapid access to data. 
Applying FL to wireless edge networks forms the so-called concept of federated edge learning or simply FEEL. FEEL involves a multi-access edge computing (MEC) \cite{filali_multi-access_2020,abouaomar_resource_2021} server that performs aggregation and edge devices that perform collaborative learning. The MEC server that is equipped with a parameter server (PS) can be a next generation nodeB (gNB), or simply a base station (BS), in a wireless edge cellular network in which there are $N$ edge devices that collaboratively train a shared model. 

Each device $\it{k}$ has a local dataset $D_k$ with a data size of $\left | D_{k} \right |$. The goal is to find the optimal global model parameters $\boldsymbol{w}\in \mathbb{R}^{l}$ that minimizes the average prediction loss $f(\boldsymbol{w})$:

\begin{equation}
    \min_{\boldsymbol{w}\in \mathbb{R}^{l}} f(\boldsymbol{w}) = \frac{1}{D} \sum_{k=1}^{N} f_k(\boldsymbol{w}), 
\end{equation}
where $\boldsymbol{w}$ is the model parameter vector to be optimized with dimension $l$, $f_k(\boldsymbol{w})$ is the loss value function computed by device $k$ based on its local training data, and $D$ is the total number of data points across all devices (i.e., $D = \sum_{k=1}^{N}\left | D_{k} \right |$). Several models' loss functions can be trained using FEEL, such as linear and logistic regression, support vector machines, and artificial neural networks.

Ideally, all the devices  independently  train  their  local  models  using their  local  training  data. Then, each one uploads its gradient updates to the server for aggregation. The server aggregates the received local updates, typically by averaging, to obtain a global model. Afterwards, the server sends the global model to the edge devices, and a new iteration begins where each device computes the gradient updates and uploads it to the server. Nonetheless, the constrained edge resources and limited communication bandwidth in wireless (edge) networks result in significant delays for FEEL. The federated averaging (FedAvg) \cite{f3} algorithm was therefore proposed to perform FEEL in a communication-efficient way. FedAvg is perhaps the most adopted communication-efficient FEEL algorithm. The main idea behind FedAvg is to select a small subset of devices and to run \if $E$\fi local epochs, in parallel, using stochastic gradient descent (SGD) on the local datasets of the selected devices. Next, all devices' resulting model updates are averaged to obtain the global model. In contrast to the naive application of SGD, which requires sending updates very often \if after every pass\fi, FedAvg performs more local computation and less frequent communication updates. Since FedAvg assumes synchronous updates collection, this may result in large communication delays. In fact, the computation, storage, and communication capabilities among participating devices might be very different. Further, devices are frequently offline or unavailable either due to low battery levels, or because their resources are fully or partially used by other applications. Thus, due to the synchronous nature of FedAvg, straggler devices, i.e., devices with low performances, will cause large delays to the whole learning process. 

Despite the promising theoretical results attained using FedAvg, deploying it or other FEEL algorithms in wireless edge networks is still not clear and challenging due to the fast changing nature of the network, limited resources and statistical heterogeneity. Statistical heterogeneity is a very important aspect in FEEL. In fact, most use cases of FEEL suppose that the system does not have control over participating devices and their training data. Furthermore, data distributions in user equipment depend heavily on the users' behaviour. As a result, any particular user’s local dataset will not be representative of the population distribution. Additionally, the datasets are massively distributed, statistically heterogeneous, i.e., non-independent and identically distributed (non-IID) and unbalanced, and highly redundant. Moreover, the raw generated data is often privacy-sensitive as it can reveal personal and confidential information. 

{\revision The wireless edge environment is composed of heterogeneous and limited capabilities of the devices \cite{reviewer1_3}}, as well as of heterogeneous data distributions. As a result, many new considerations related to communication resources and edge devices' data should be reflected in the design and deployment of FEEL algorithms. This motivates several works in FEEL to adapt to the communication-restrained edge environment.

% First, despite the computation, storage and communication capabilities among the participating devices, FedAvg assumes synchronous updates collection. The gap in computational resources due to different devices capabilities creates challenges such as delays caused by stragglers, i.e., devices with low performances. Second, devices are frequently offline or unavailable either due to low battery levels, or because their resources are fully or partially used by other applications. 

% Furthermore, data distributions in user equipment depend heavily on the users' behaviour. As a result, any particular user’s local dataset will not be representative of the population distribution. Additionally, the datasets are massively distributed, statistically heterogeneous (i.e., non-IID and unbalanced), and highly redundant. Moreover, the raw generated data is often privacy-sensitive as it can reveal personal and confidential information.

% Despite the promising theoretical results attained using FedAvg, deploying FEEL is challenging due to the fast changing nature of the environment, mainly resource limitations and statistical heterogeneity. 
% This motivates several works in FEEL to adapt to the communication-restrained edge environment.

\subsection{Related Work}
Several prior works investigated the communication-constrained FEEL systems from different perspectives. In addition to several compression  \cite{deepcompression} and partial participation \cite{konecny_federated_2017} techniques, several mechanisms were proposed to adapt the federated training to the constrained resources. For instance, authors in \cite{wang_adaptive_2019} propose an accommodation mechanism where the numbers of global and local iterations are changed depending on the available communication and computation resources. Another suggested approach relays on collecting the updates in an asynchronous manner \cite{xie_asynchronous_2019}, which allows a smooth adaptation to heterogeneous resources and a flexible updates collection. However, due to stale updates' effect on learning, synchronous updates collection remains the preferred method. As a result, a common approach is to selectively schedule a subset of devices to send their updates in each communication round. 
For example, authors in \cite{fedcs} proposed a client selection algorithm to reduce the latency of the model training, where only the end devices with good communication and computation capabilities are chosen, thus avoiding the straggler’s problem. Nevertheless, this method is biased toward powerful devices with better channel states, which discards devices with potentially more informative or important updates, and might lead to models that cannot generalize to a wide range of devices. To diversify the sources of updates, several works adopted scheduling algorithms that aim to maximize the number of participating devices. For instance, authors in \cite{zeng_energy-efficient_2019}  proposed an energy-efficient joint bandwidth allocation and scheduling algorithm, which ensures the training speed by collecting the maximum amount of updates possible. Nonetheless, this method does not guarantee the diversity of updates sources. Consequently, fairness measures \cite{yang_age-based_2020, yang_scheduling_2019} were adopted in scheduling policies to ensure gradient diversity. For instance, {\revision an age-based scheduling (ABS) algorithm} was proposed in \cite{yang_age-based_2020}, where  higher priority is given to devices that were not selected in several previous rounds.  However, relying on strict fairness-based policy may yield a low number of collected updates within a round.

Another approach for selective scheduling relies on evaluating the resulting model in an attempt to reduce the number of collected updates by removing the irrelevant ones \cite{wang_cmfl_2019}. This is achieved through measuring  the  significance  of  a  local update relative to the current global model, and whether this update aligns with the collaborative convergence trend.
Nonetheless, this approach may not be energy efficient, as it is applied post-training. Computing updates is an energy consuming operation, as a result, disregarded updates are synonymous with wasted energy.

%when using incentives \cite{kang_incentive_2019} also seems unfeasible, as it can lead to higher costs and low utility. 

Despite the variety of research progresses, the resource-efficient FEEL scheduling algorithm design with highly heterogeneous dataset distributions remains a topic that is not well addressed. This motivates our work, in which we investigate a possible direction to evaluate the potential significance of the updates through local dataset characteristics, namely size and diversity.

\section{Diversity in Federated Learning} 
\label{sec:Diversity}
Our idea comes from the fact that many prior works in ML have imposed the diversity on the construction of training batches to improve the efficiency of the learning process \cite{zhang_determinantal_2017}. Furthermore, active learning \cite{ you_diverse_2014, shi_diversifying_2016} is premised on the idea that models can be trained with fewer data points, provided that the selected samples are diverse and more informative. In active learning, the diversity is used as a criterion for choosing informative data points for efficient ML training. However, to the best of our knowledge, this premise has never been used in FEEL prior to this work. Thus, we investigate the possibility of exploiting the different dataset properties to carefully select devices with potentially more informative datasets with less redundancy, by measuring their size and diversity. Therefore, we propose data-aware scheduling (DAS) algorithm for FEEL where these aspects are the heart of the devices selection.

%Generally, a diversified dataset contains more information and can train a model that can be fit for various environments. 

\begin{figure*}[t]
	\centering
	\includegraphics[scale=0.67]{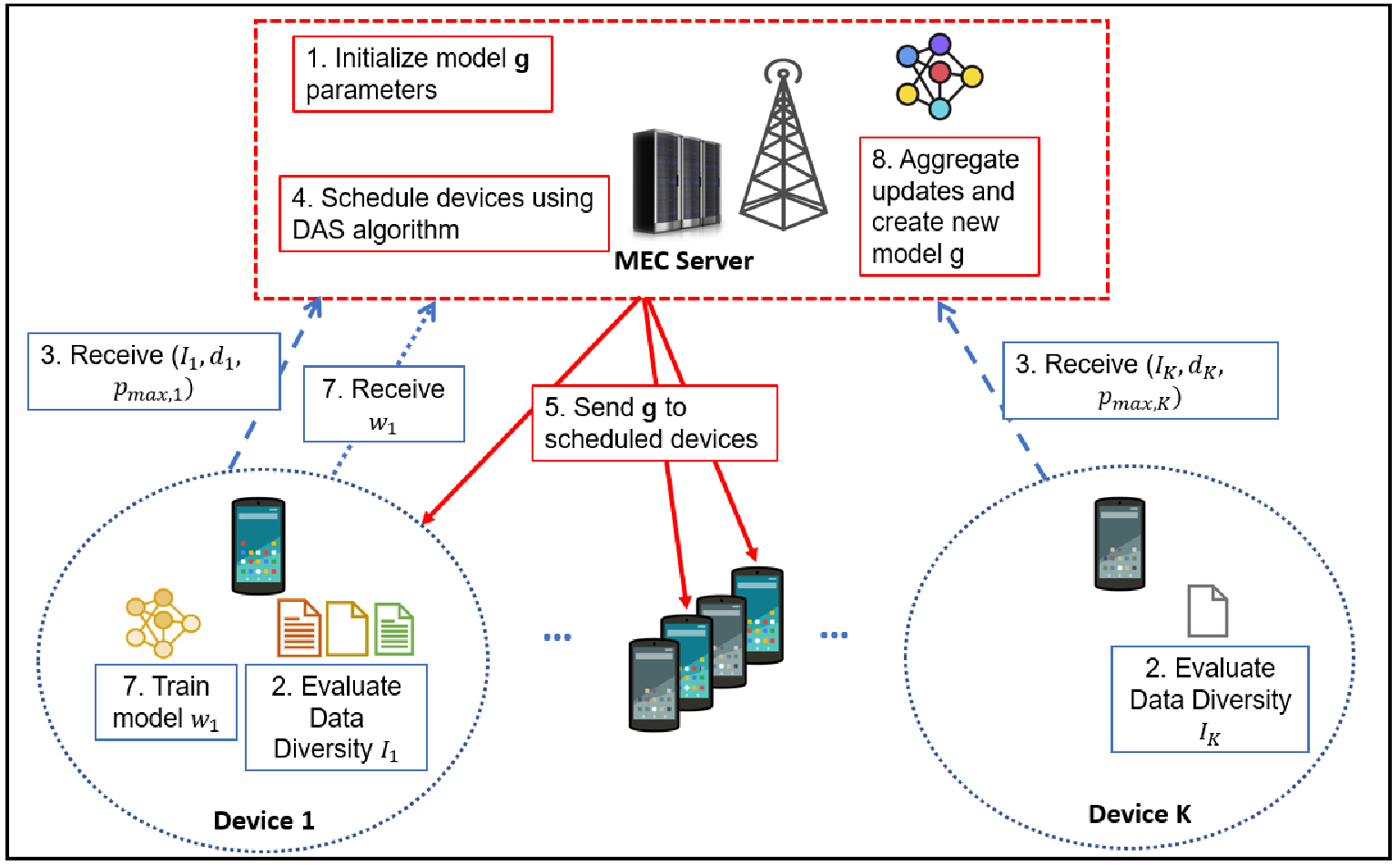}
	\caption{{\revision The FEEL system model. Based on the received devices' information, a subset of devices are scheduled for training the model and uploading their updates.}}
	\label{fig:fig_sys-model}
\end{figure*}

The first question to be asked is what would be a good diversity measure for FEEL? {\revision Various measures of diversity are used in active learning to choose the samples that should be labeled. For instance, in an image classification problem, images that are hard to classify with high certainty are considered more informative and are selected to be labelled, and the chosen samples are selected from different classes in order to form a diverse dataset \cite{yang_multi-class_2015}. 
In FEEL, the device selection does not concern independent samples, and the dataset construction is not possible due to on-device processing. As a result, the diversity must be evaluated at the level of the entire dataset.  Moreover, in the premise of FEEL, the labels are already known which gives the possibility to use more informed measures.  For instance,  we can use Shannon entropy \cite{holub_entropy-based_2008} or Gini-Simpson index \cite{jost_entropy_2006} for classification problems, to see whether a dataset has diverse examples from several classes. If a classification dataset has samples from one class only, it should be discarded during training. 
Additionally, sequence prediction models are sensitive to data-quality, especially missing data and null values \cite{nespoli_data_2020, arbesser_visplause_2017,taik_electrical_2020}. Consequently, the diversity (i.e., variations of the sequence data) should be evaluated using methods such as approximate entropy (ApEn) and sample entropy (SampEn) \cite{Apen}.}

\vspace{0.2 cm}
%\subsubsection{Diversity measures}
The Gini-simpson index is a modification of the Simpson index. The Simpson index measures the probability that two samples taken at random from the dataset of interest are from the same class, it is calculated as follows:
\begin{equation}
    \lambda = \sum_{c=1}^{C} p_{c}^{2},
\end{equation}
where $C$ is the total number of classes, and $p_{c}$ is the probability of the class $c$.
The original Simpson index $\lambda$ represents the probability that two samples taken at random from the dataset are of the same type (i.e., are within the same class). The Gini–Simpson index is its transformation $1-\lambda$, which represents the probability that the two samples belong to different classes.
The Gini-Simpson index is used in different applications such as financial markets \cite{jiang_permutation_2017} and analyzing ECG signal \cite{khorasani_investigation_2019}.
\\
The Shannon entropy  also quantifies the uncertainty of a prediction, and was used in several applications such as text prediction and image classification \cite{liu_data-importance_2019}. In the context of FEEL, it can be used as follows: 
\begin{equation}
    H = -\sum_{c=1}^{C}p_{c}\log_2(p_{c}),
\end{equation}
where $C$ is the total number of classes, and $p_{c}$ is the probability of the class $c$. 
Shannon Entropy is not defined for the extreme case of 0 samples in a class, which can be problematic in some highly unbalanced classification problems. 

%\subsubsection{Diversity measures for Sequential data}
In sequence data, statistical measures such as the mean and the variance are not enough to illustrate the regularity, as  they are influenced by system noise. ApEn was proposed to quantify the amount of regularity and the unpredictability of time-series data \cite{catt_forecastability_2009}. 
It is based on the comparison between values of data in successive vectors, by quantifying how many successive data points vary more than a defined threshold. A random time series with fewer data points can have a lower ApEn than a more regular time series, whereas, a longer random time series will have a higher ApEn. SampEn \cite{montesinos_use_2018} was proposed as a modification of  ApEn. It can be used for assessing the complexity of time-series data, with the advantage of being independent from the length of the vectors. Both these measures can help eliminate outliers, however, it should be noted that measuring ApEn and SampEn is a computationally heavy task, therefore it should be evaluated on a small sample rather than the entire dataset.

To sum up, several diversity measures can be applied on datasets for different applications. Dataset diversity will allow a more informed participant selection in FEEL. Choosing devices with diversified datasets can accelerate the training and avoid overfitting, as the datasets contain more information.

\section{System Model and Problem Formulation}
\label{sec:sysmodel}
Having discussed several diversity measures that can be used in various FEEL applications, let us introduce how such diversity measures can be used in the overall design of the FEEL algorithm. Hereinafter, we consider a FEEL system with multi-device and a single MEC server.  The system model is illustrated in Fig.~\ref{fig:fig_sys-model}. In this section, we introduce the different elements of the system model, then we formulate a joint device selection and bandwidth allocation problem.

To study the suitability of the proposed selection criteria in the context of FEEL, we consider a wireless edge network composed of a MEC server and $\it{K}$ devices collaboratively training a shared model. Each device $k$ is characterized by a local dataset $D_k$ with a data size $\left| D_{k} \right|$.

\subsection{Learning Model}
First, the global model's architecture and weights are initialized by the MEC server. At the beginning of each training round $\it{r}$, the devices send their information and dataset diversity indicators to the MEC server. 
Based on the received information, alongside with the evaluated channel state information, the server selects a subset $S_r$ of the devices and allocates the necessary bandwidth to each scheduled device in order to receive the global model $\it{g}$.  The scheduling of the devices, presented in Algorithm \ref{alg:feel}, is based on the trade-off between the datasets diversity and the required time and energy, under the constraint of a minimum number of devices that should be scheduled in each round. In fact, given synchronous aggregation, the MEC server requires a minimum number $K$ of updates to be collected to consider a round complete. Then, each device $\it{k}$ in the chosen subset $S_r$ uses $\left | D_{k} \right |$ examples from its local dataset. SGD is then used by each device $\it{k}$ to compute its local update for some period of $E$ local epochs. The updated models $w_{k}$ are sent to the MEC server for aggregation. Ideally, all devices transmit their trained local models to the MEC server simultaneously. The FEEL process is repeated over $r_{max}$ communication rounds and we use $D_r=\sum_{k \in S_r} \left | D_{k} \right |$ to denote the total size of the datasets of all selected devices.

In order to aggregate the client updates, the MEC server uses the weighted average technique of the \textit{FedAvg} algorithm proposed in~\cite{f3}. %Nonetheless, due to communication bottleneck, only a subset can be scheduled in each round. 
The MEC server aggregates the updates and sends the resulting parameters to a new subset of selected devices. This process is repeated until the desired prediction accuracy is reached or a maximum number of rounds is attained. In the considered architecture, the considered FEEL procedure is detailed in Algorithm \ref{alg:algoFEEL}. 

\begin{algorithm}[htb!]
% 	\SetAlgoLined
	%\KwResult{Write here the result }
	\begin{algorithmic}[1]
	\While{$\it{r}<r_{max}$ or $accuracy < \text{desired accuracy}$}
	    \If{$r=0$}
	        \State initialize the model's parameters at the MEC server
	    \EndIf
		\State Receive devices information (transmit power, available data size, dataset diversity index)
		\State Schedule a subset $S_r$  of devices with at least $\it{N}$ devices using Algorithm \ref{alg:feel}
		\For {\text{ device } $k \in S_r$ }
			 \State $\it{k}$ receives model $\it{g}$
			 \State $\it{k}$ trains on local data $D_{k}$ for $E$ epochs
			\State $\it{k}$ sends updated model $w_k$ to MEC server
		\EndFor
		\State MEC server computes new global model using weighted average:
		 $g\leftarrow \sum_{k \in S_r }\frac{D_{k}}{D_r}w_{k}$
        \State start next round $\it{r}\leftarrow \it{r+1}$		
	\EndWhile
	\end{algorithmic}
	\caption{FEEL Procedure}
	\label{alg:algoFEEL}
\end{algorithm}
\subsection{Dataset Diversity Index Design}  
\label{subsec:index}
%\subsubsection{Diversity Index Design}
Due to the unbalance and non-IID nature of the distributions, and under high bandwidth constraints, the dataset size and diversity need to be considered in the device selection. 

Additionally, we consider a second aspect which can be viewed at the system level which is the  diversity of sources.  This goal can be achieved through maximizing the fairness {\revision in terms of the number of updates collected from each device} is used to guarantee the diversity of the data sources.

Therefore, the goals of the devices' selection are twofold: 
1) select devices with potentially informative datasets, which is achieved through evaluating the size and  diversity of the datasets; and
2) guarantee that the selected devices are diversified, which will be attained by adding age-of-update to the designed diversity index.

Since our scheduling problem can consider multiple criteria, namely dataset diversity and fairness of the selection, and each measure is calculated with a function that has an output on different scales, the function should be designed to output a weighted rank value bounded in $[0, \gamma_i]$, where $i \in \{\text{dataset diversity}, \text{dataset size}, \text{age}\}$.  %It is worthwhile to note that 
The value of this function is given as follows: $ v_i \times \gamma_i$,  where  $\gamma_i$ is the adjustable weight for each metric assigned by the server and $v_i$ is the normalized value of the metric $\it{i}$ calculated as follows:
\[v_i =\frac{ \text{measured value of metric}\; i }{ \text{maximum for metric}\; i}.\] 
We define the diversity index of dataset $k$ as: 
\begin{equation}
I_k = \sum_{i} v_{i,k}\gamma_{i,k},
\end{equation}
where $v_{i,k}$ is calculated for specific dataset $k$.

Note that this measure is in line with federated learning principles, as it can be evaluated on-device, and it does not reveal any privacy-sensitive information about the dataset. {\revision If fact, by combining several measures into one weighted index, it is hard for an eavesdropper to extract information about the dataset's raw elements if they intercept the index.}      

We formulate the first goal of the device selection problem as:
\begin{equation}
    \max_{x} \sum_{k=1}^{K} I_k x_k, \label{eq:goal1}
\end{equation}
where $x=[x_1,...,x_K]$ and $x_k$, for $k=1,2,\ldots,K$, is a binary variable that indicates whether or not device $k$ is scheduled to send an update, and $I_k$ is the diversity index. 
%It is worthwhile to note that in the special case of i.i.d and balanced 
\subsection{Transmission Model}
As presented in Algorithm \ref{alg:algoFEEL}, the transmission aspect is also considered during the devices scheduling. Given that the bandwidth is the bottleneck of FEEL, it is essential to estimate the required transmission time and energy during the scheduling, as a means to avoid stragglers and device drop out due to low energy.  
Henceforth, we consider orthogonal frequency-division multiple access (OFDMA) for local model uploading from the devices to the MEC server, with total available bandwidth of $\it{B}$ Hz. We define $\alpha = [\alpha_1,...,\alpha_K]$, where for each device $\it{k}$,  $\alpha_{k} \in [0,1]$ is the bandwidth allocation ratio. The channel gain between device $k$ and the BS is denoted by $h_{k}$. Due to limited bandwidth of the system, the bandwidth allocation ration should respect the constraints $\sum_{k=1}^{K} \alpha_{k} \leq 1$.
The achievable rate of device $\it{k}$ when transmitting to the BS is given by: 
\begin{equation} \label{eq:eqRate}
r_k=\alpha_{k} B \log_{2}(1 +\frac{g_{k} P_{k}}{\alpha_k B N_0}),\qquad \forall k \in [1,K],
\end{equation}
where $P_{k}$ is the transmit power of device $\it{k}$, and $N_{0}$ is the power spectral density of the Gaussian noise.
Based on the synchronous aggregation assumption, the duration of a communication round depends on the last scheduled device to finish uploading. The round duration is therefore given by: 
\begin{equation} \label{eq:eqTime}
 T = \max ((t_{k}^{train}+t_{k}^{up}) x_k ) ,\qquad   \forall k \in [1,K],
\end{equation}
where $t_{k}^{train}$ and $t_{k}^{up}$ are, respectively, the training time and transmission time of device $\it{k}$. The training time $t_{k}^{train}$ depends on device $k$'s dataset properties as well as on the model to be trained. It can be estimated using Eq.\ref{eq:ttrain}: 
\begin{equation}
    t_{k}^{train} = E \left | D_k \right | \frac{C_k}{f_k},
    \label{eq:ttrain}
\end{equation}
where $C_k$(cycles/bit) is the number of CPU cycles required for computing one sample data at device $k$ and $f_k$  is its computation capacity.
To send an update of size $\it{s}$ within a transmission time of $t_{k}^{up}$, we must have:
\begin{equation}
t_{k}^{up} = \frac{s}{r_k}.    
\end{equation}

Finally, the wireless transmit energy of device $\it{k}$ is given by:
\begin{equation}
E_{k}=P_k t_{k}^{up}.
\end{equation}

\subsection{Device Scheduling Problem Formulation} 

Considering the collaborative aspect of FEEL, and the communication bottleneck, we define the following goals for the device scheduling algorithm: 
\begin{itemize}
    \item From the perspective of devices, it is desirable to consume the least amount of energy to carry the training and uploading tasks. Given that the participating devices are responsible for the training, the aspect that can be adjusted is the upload energy. Therefore, the first goal is to minimize the consumed upload energy of the scheduled devices : 
    \begin{equation}
        \min_{x,\alpha} \sum_{k=1}^{K}{x_{k}E_{k}}.
    \end{equation}
    \item Due to the heterogeneous device capabilities and the largely varying data sizes, it is hard to estimate a suitable deadline for each round. To avoid stragglers' problem, it is desirable for the MEC server to have short round duration. Thus, a part of the objective is the minimization of the communication round.
    \begin{equation}
        \min_{x,\alpha} T.
    \end{equation}
    \item From the perspective of accelerating learning, we adopt the goal defined in Subsection \ref{subsec:index} in Eq \ref{eq:goal1}.
\end{itemize}

By combining these three goals, the problem is formulated as a multi-objective optimization problem as follows: 
\begin{mini!}|l|[3]
 {x,\alpha}{\left\{{\sum_{k=1}^{K}{x_{k}E_{k}}}, T -\sum_{k=1}^{K}{x_{k}I_{k}}\right\}}
 {}{} \label{pbmoo}
\addConstraint{\qquad (t_{k}^{train}+t_{k}^{up}) x_k \leq T, \qquad \forall k \in [1,K]}\label{eq:TrainUploadT}
\addConstraint{\qquad \sum_{k=1}^{K}\alpha_k  \leq 1,\qquad  \forall k \in [1,K]}\label{eq:Bandwidth}
\addConstraint{\qquad  0\leq \alpha_k \leq 1,\qquad \forall k \in [1,K]}\label{eq:alpha_bounds}
\addConstraint{\qquad \sum_{k=1}^{K} x_k  \geq N,\qquad  \forall k \in [1,K]}\label{eq:minrequired}
\addConstraint{\qquad x_k \in \{0,1\}, \qquad \forall k \in [1,K].}\label{eq:x_bounds}
\end{mini!}

where the constraints are defined as follows. Constraint (\ref{eq:TrainUploadT}) guarantees that the scheduled devices finish training and uploading the models before the deadline. Constraints (\ref{eq:alpha_bounds}) and (\ref{eq:Bandwidth}) ensure that the allocated bandwidth fractions are between 0 and 1 and that their sum does not exceed the bandwidth budget. Constraint (\ref{eq:minrequired}) guarantees that the number of selected devices is at least equal to the minimum required, with constraint (\ref{eq:x_bounds}) setting $x_k$ as a binary variable.

%%%%%%%%%%%%%%%%%%%%%%%%%%%%%%%%%%%%%%%%%%%%%%%%%%%%%%%%%%%%%%
\subsection{NP-hardness}
Problem (13) is a multi-objective problem that is non-linear. Thus, it is very challenging to solve. Even worse, we show in the following that the problem is NP-hard even for a single objective case. {\revision Indeed, a restricted version of problem~(13) is shown to be equivalent to a knapsack problem and thus it is NP-hard. The main difficulty of the problem comes from maximizing the weighted sum of the devices while allocating the bandwidth. In this regard, when we fix the transmit power and assume that each device is allocated a certain amount of bandwidth, then the multi-objective function in (13a) reduces to, for fixed transmit powers $p_k$ and fixed $\alpha_k$, the problem of maximizing the weighted number of devices, i.e., $\sum_{k}I_kx_k$ subject to a knapsack capacity given by $\sum_{k}\alpha_kx_k\leq1$. Constraint~\eqref{eq:TrainUploadT} can be verified for each device to filter out the devices that do not respect them since it depends only on the fixed power.} Thus, the problem is equivalent to a knapsack problem and since the latter is NP-hard, so is problem~(13).
%%%%%%%%%%%%%%%%%%%%%%%%%%%%%%%%%%%%%%%%%%%%%%%%%%%%%%%%%%%%%%%

\section{Scheduling algorithm}
\label{sec:algo}
{\revision In this section, we present our data-aware FEEL scheduling algorithm to solve the multi-objective problem (13) defined in section \ref{sec:sysmodel}. Besides being NP-hard, problem (13) is a mixed integer non-linear multi-objective program that is hard to solve. The proposed FEEL scheduling algorithm to solve (13) proceeds in two main steps as follows. First, problem (13) is decomposed into two subproblems. Next, the proposed algorithm optimizes iteratively both subproblems.}

%In this section we present our data-aware FEEL scheduling algorithm to optimize the multi-objective problem defined in Section \ref{sec:sysmodel}. The defined multi-objective problem is mixed integer non-linear program. To efficiently solve it, we decompose it into two sub-problems, and we discuss each one in the sequel.

The first sub-problem (\textbf{Sub1}) is a selection problem in which we select the devices in order to optimize a weighted linear combination of the different objectives. The selection sub-problem is formulated as follows:  
 \begin{mini!}|l|[3]
 {x}{{\lambda_E \sum_{k=1}^{K}{x_{k}E_{k}}}+\quad \lambda_T T \quad - \lambda_I \sum_{k=1}^{K}{x_{k}I_{k}}}
 {}{} 
 \addConstraint{ \qquad x_k \in \{0,1\}  \qquad \forall k \in [1,K]}\label{eq:x_bounds2} \qquad  
 \addConstraint{ \qquad \sum_{k=1}^{K} x_k \geq K  \qquad \forall k \in [1,K]}\label{eq:x_minimum} \qquad
 \end{mini!}
 where $\lambda_E$, $\lambda_T$, and $\lambda_I$ are positive scaling constants used first to scale the value of the objective function and second to combine the different conflicting objectives into a linear single one. 
 
The second sub-problem (\textbf{Sub2}) is a bandwidth allocation problem in which the device selection decision is fixed through solving the previous selection sub-problem. The objective of the bandwidth allocation problem consists of linear combination, using a positive constant $\rho$, of the consumed energy and the round's completion time. This problem is formulated as follows:
 \begin{mini!}|l|[4]
 {\alpha}{{\rho \sum_{k=1}^{K}{x_{k}E_{k}}} +\quad (1-\rho) T  }
 {}{}
 \addConstraint{ \qquad \sum_{k=1}^{K}\alpha_k \leq 1,\qquad  \forall k \in [1,K]}\label{eq:Bandwidth2}
 \addConstraint{\qquad  0\leq \alpha_k \leq 1,\qquad \forall k \in [1,K]}\label{eq:alpha_bounds2}  \qquad
 \end{mini!}
 
 To solve \textbf{Sub1}, we use \textit{relaxation and rounding}. Specifically, we relax the integer constraint $x_k \in \{0,1\}$ as the real-value constraint $0 \leq x_k \leq 1$, and then the integer solution is determined using rounding after solving the relaxed problem, then verifying whether the condition (\ref{eq:x_minimum}) is satisfied. The relaxed problem can be written as: 
 \begin{mini!}|l|[3]
 {x}{{\lambda_E \sum_{k=1}^{K}{x_{k}E_{k}}}+\quad\lambda_T T \quad - \lambda_I \sum_{k=1}^{K}{x_{k}I_{k}}}
 {}{} 
 \addConstraint{ \qquad 0 \leq x_k \leq 1  \qquad \forall k \in [1,K]}\label{eq:x1_bounds} \qquad 
 \end{mini!}

The continuous value of $x_k$ can be viewed as the selection priority of the device $k$, therefore, if the condition (\ref{eq:x_minimum}) is not satisfied, we set $x_k = 1$ for the $N$ devices with highest priorities.
To solve \textbf{Sub2}, we applied off-the-shelf solvers in order to obtain the optimal bandwidth allocation ratios.

The proposed FEEL algorithm is an iterative algorithm that solves each sub-problem sequentially as discussed previously and updates the solution in each iteration. The pseudo-code of the data-aware scheduling (DAS) algorithm is given in Algorithm~\ref{alg:feel}. The algorithm iterates until convergence or until a maximum number of iterations is reached, whichever appears first. The convergence happens when the values of $x$ and $\alpha$ does not undergo a large change, e.g., when their values are almost the same as in the previous iteration, the loop is terminated. The number of iterations is set to $iterations_{max}$ and is used to guarantee the algorithm termination. 

 \begin{algorithm}[h]
% 	\SetAlgoLined
	%\KwResult{Write here the result }
	\begin{algorithmic}[1]
	\State initialize  $x_k=1 \forall k \in[1,K]$
	\State uniformly allocate the bandwidth
	\State $iterations\gets1$
	\While{$\it{iterations}<iterations_{max}$ \textbf{and not} convergence}
		\State \textbf{Solve Sub1 :}  Return $x$
		\State \textbf{Round} $x$
		\If{\text{condition(\ref{eq:x_minimum}) is satisfied }}           \State continue
		    \State select $K$ \text{devices with the highest priorities}
		\EndIf
		\State \textbf{Solve Sub2 :}  Return $\alpha$ 
		\State $iterations\gets iterations+1$
	\EndWhile
% 	\State \textbf{Until Convergence}
	\end{algorithmic}
	\caption{DAS algorithm for FEEL}
	\label{alg:feel}
\end{algorithm}

\section{Simulation and results}
\label{sec:simulation and results}
In this section, we present the performance evaluation of the DAS algorithm. 
{\revision The  experiments  we conducted consist  of  two  parts: 1) an evaluation for the proposed diversity index; and 2) an evaluation of the performance of the proposed DAS algorithm.}

{\revision We first present in Section \ref{subsec:param1} the simulation environment and parameters including the wireless edge environment and the datasets and trained models. Next, we evaluate our proposed diversity index in Section \ref{subsec:index1}. We compare the diversity index' impact in the selection of participant devices in comparison to random selection. Then, we evaluate DAS algorithm under different settings by comparing it to other scheduling approaches in Sections \ref{subsec:size1} and \ref{subsec:comp1}. As communication is the bottleneck of FEEL, sending large models over wireless networks limits the number of participants. Thus, we mainly evaluate the performance of DAS algorithm by varying the model size in Section \ref{subsec:size1}. Furthermore, since trading communication rounds with more local computation is one of the key enablers of FEEL, we also study the effect of varying the number of local epochs in Section \ref{subsec:comp1}.} 

In the experiments, in order to evaluate the possible gain from using our proposed algorithm, we compare DAS to two scheduling strategies: 1) a baseline scheduling where all the devices participate in the training with optimized time and energy following problem \textbf{Sub2}. We compare to this baseline scheduling in order to evaluate the scalability of the algorithm in terms of the consumed energy and time. 2) an age-of-update based scheduling (ABS) algorithm \cite{yang_scheduling_2019, yang_age-based_2020}, specifically we used the age-based priority function proposed in \cite{yang_scheduling_2019} with $\alpha = 1$, $f(k) = \log(1 + T(k))$  with $T(k)$  the number of rounds since last selection of device $k$. This algorithm considers both the variety of participants. {\revision In fact, using ABS, the devices that did not participate in the past few rounds have more priority. Therefore, by comparing DAS to ABS, we are able to measure the importance of the dataset diversity in the algorithm, in contrast to ABS approach that only considers the diversity of participants.}
 
\subsection{{\revision Simulation Environment and Parameters}}\label{subsec:param1}
The simulations were conducted on a desktop computer with a 2,6 GHz Intel i7 processor and 16 GB of memory and NVIDIA GeForce RTX 2070 Super graphic card. We used Pytorch \cite{noauthor_pytorch_nodate} for the machine learning library, and Scipy Optimize \cite{noauthor_optimization_nodate} for the optimization modeling and solver. In the numerical results, each presented value is the average of 50 independent runs.

\textit{{\revision 1) Wireless Edge Environment:}}\\
We consider a cellular network modelled as a square of side $500$ meters and composed of one BS located in the center of the square. The $K$ edge devices are randomly deployed inside the square following uniform distribution. Unless  specified  otherwise, the  simulation  parameters  are  as  follows.  We consider $K= 100$ edge devices, and $N = 1$ the minimum number of devices to be scheduled. The OFDMA bandwidth is $B = 1$ MHz.  The  channel gains $g_k$ between edge device $k$ and the BS includes large-scale pathloss and small-scale fading following Rayleigh distribution, i.e., $|g_k|^2=d_k^{-\alpha}|h_k|^2$ where $h_k$ is a Rayleigh random variable and $\alpha$ is the pathloss exponent and $d_k$ is the distance between edge device $k$ and the BS. We set the parameters of problem \textbf{Sub1} as $\lambda_E = \lambda_T = \frac{1}{4}, \lambda_I = \frac{1}{2}$ and the parameters of \textbf{Sub2} as $\rho = \frac{1}{2}$. The remainder of the used parameters are summarized in Table I. 

\begin{table}[h]
     	\centering{
     		\caption{Generated Values}
     		\begin{tabular}{|c|c|c|c|c|r|r|r|r|r|}
     			
     			\hline
     		
     			Devices CPU frequency  & [ 1,3 ] Ghz     \\ \hline
     			Cycle /bit & [ 10,30 ] (cycles/bit)    \\ \hline
     			Transmit Power & [1,5]     \\ \hline
     			Model Size & 100 kbits     \\ \hline
     			Bandwidth & 1MHz    \\ \hline
     			Number of shards per device & [1,30]    \\ \hline
     		\end{tabular}
     	}
     	
         	\label{tab:tab1}
     \end{table}

\textit{{\revision 2) Dataset and Trained Models:}}\\
\textbf{{\revision Dataset:}} We used benchmark image classification dataset MNIST \cite{mnist}, which we distribute randomly among the simulated devices. The data distribution we adopted is as follows: 
We first sort the data by digit label, then we form 1200 shards composed of 50 images each. Each shard is composed of images from one class, i.e. images of the same digit. {\revision In the beginning of every simulation run, we randomly allocate a minimum of 1 shard and a maximum of 30 shards to each of the 100 devices considered in this simulation.} This method of allocation allows us to create an unbalanced and non-IID distribution of the dataset. We keep 10\% of the distributed data for test, and use the remaining for training.

\textbf{{\revision Models:}}  {\revision convolutional neural networks (CNNs) and multi-layer perceptrons (MLPs) are widely used in classification problems. More specifically, MLPs are used for classification problems in general, while CNNs are especially powerful for image classification as they allow scale independence. We used these two different models in order to show the effect of the proposed algorithm in different models’ convergence.} The CNN model is given with two 5x5 convolution layers (the first with 10 channels, the second with 20, each followed with 2x2 max pooling), two fully connected layers with 50 units and ReLu activation, and a final softmax output layer. We also train a simpler MLP model with two fully connected layers. Since our goal through using these models is to evaluate our scheduling algorithm and not to achieve state-of-the-art accuracy on MNIST, therefore they are sufficient for our goal. Furthermore, the selected models are fairly small, thus they can be realistically trained on resource-constrained and legacy devices, using reasonable amounts of  energy in short time windows. 

We first evaluate the diversity index through several experiments. Then we evaluate DAS over wireless networks by varying the model size and the number of local epochs.

%number of parameters : 21840
\begin{figure}[htb]

	\begin{subfigure}[t]{0.35\textwidth}
		\includegraphics[width=1.5\linewidth]{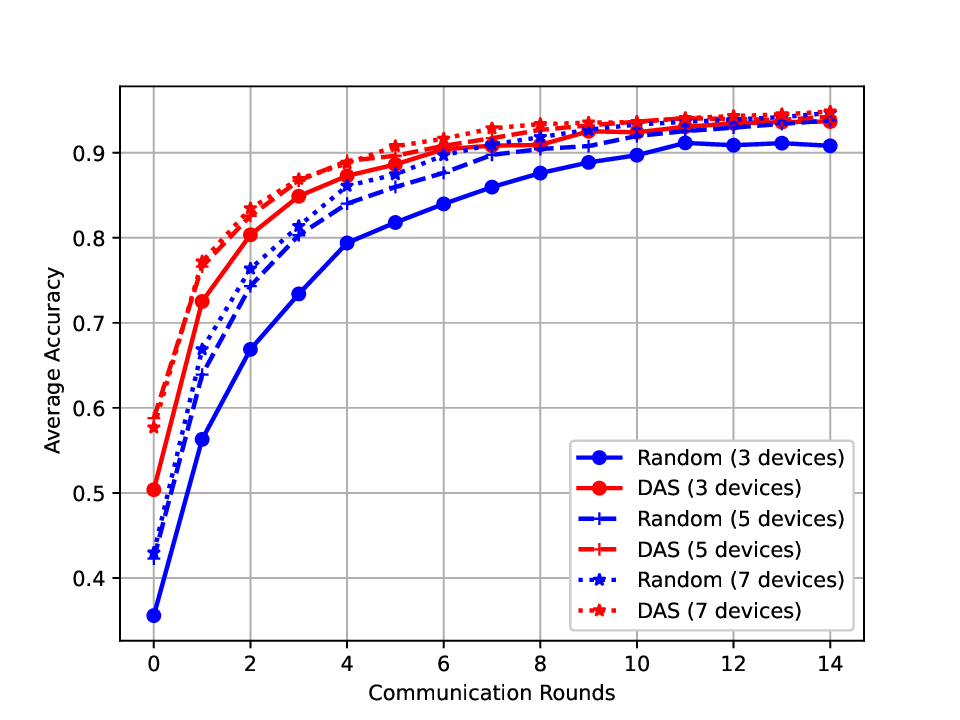}
		\caption{CNN}
		\label{cnn_clients}
	\end{subfigure} \hspace{18mm}
	\begin{subfigure}[t]{0.35\textwidth}
		\includegraphics[width=1.5\linewidth]{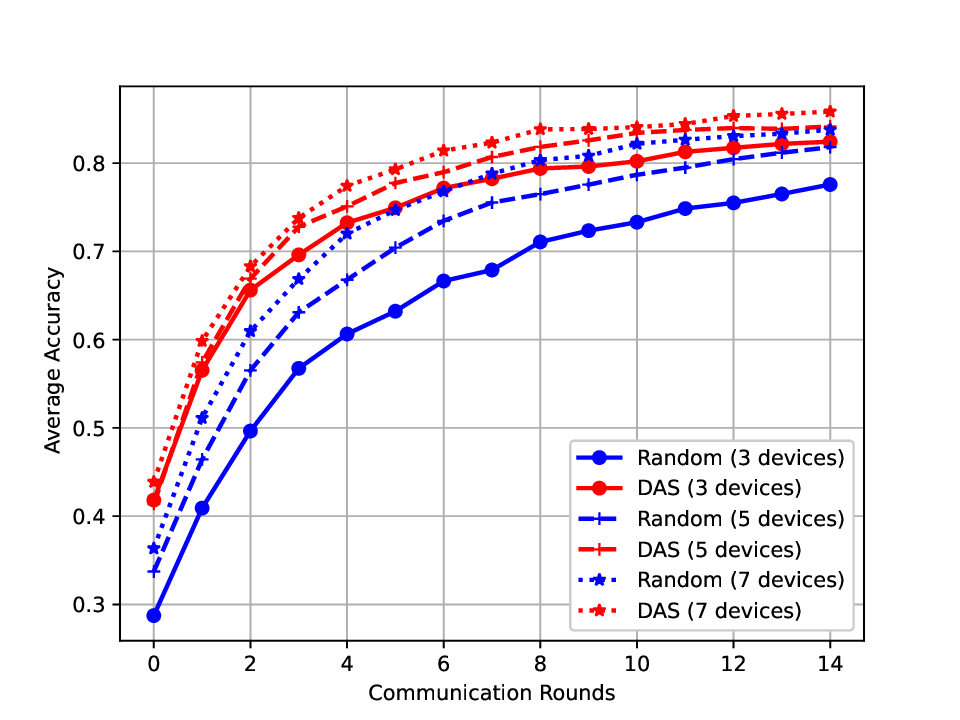}
		\caption{MLP}
		\label{mlp_clients}
	\end{subfigure} 

	\caption{ {\revision Average test accuracy by varying number of selected devices} }
	\label{cnn_mlp_clients} 
\end{figure}

\begin{figure}[htb]

	\begin{subfigure}[t]{0.35\textwidth}
		\includegraphics[width=1.5\linewidth]{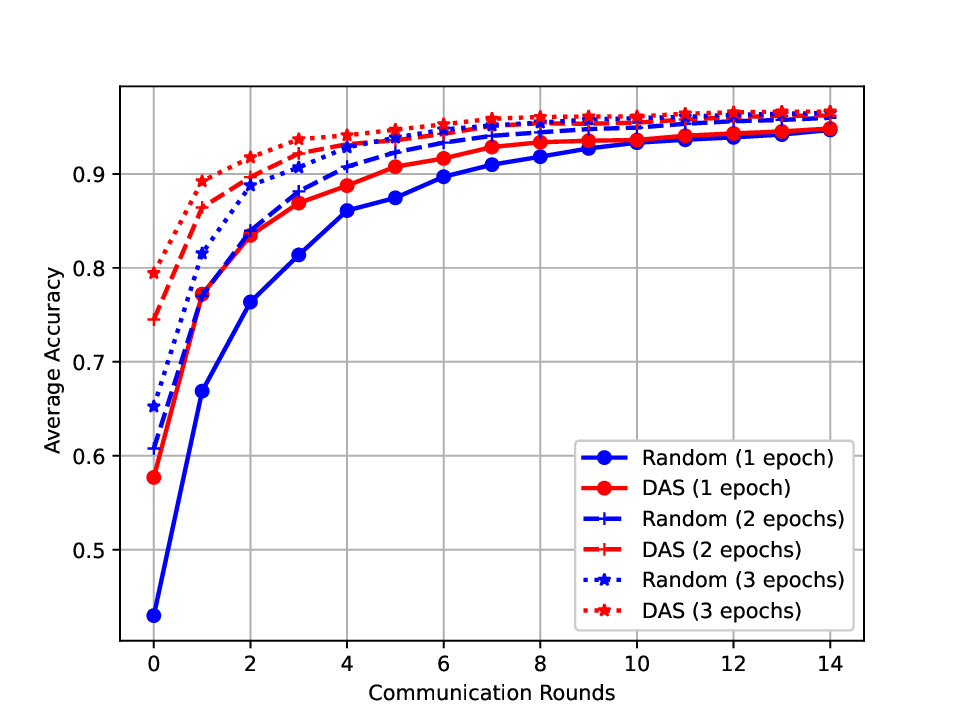}
		\caption{CNN}
		\label{cnn_clients}
	\end{subfigure} \hspace{18mm}
	\begin{subfigure}[t]{0.35\textwidth}
		\includegraphics[width=1.4\linewidth]{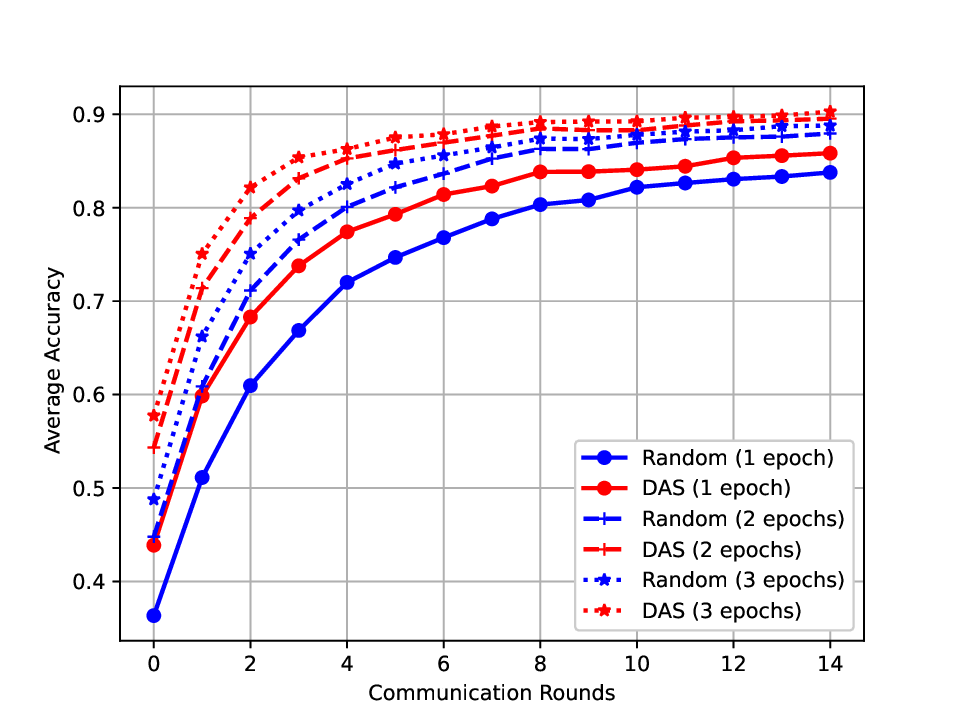}
		\caption{MLP}
		\label{mlp_clients}
	\end{subfigure} 

\caption{{\revision Average test accuracy by varying number of local epochs with fixed number of devices}}
	\label{cnn_mlp_epochs}
\end{figure}

\begin{figure}[htb]
	%\centering
	%\begin{center}
	%\hspace{0.07 in}
	\begin{subfigure}[t]{0.3\textwidth}
		\includegraphics[width=1.4\linewidth]{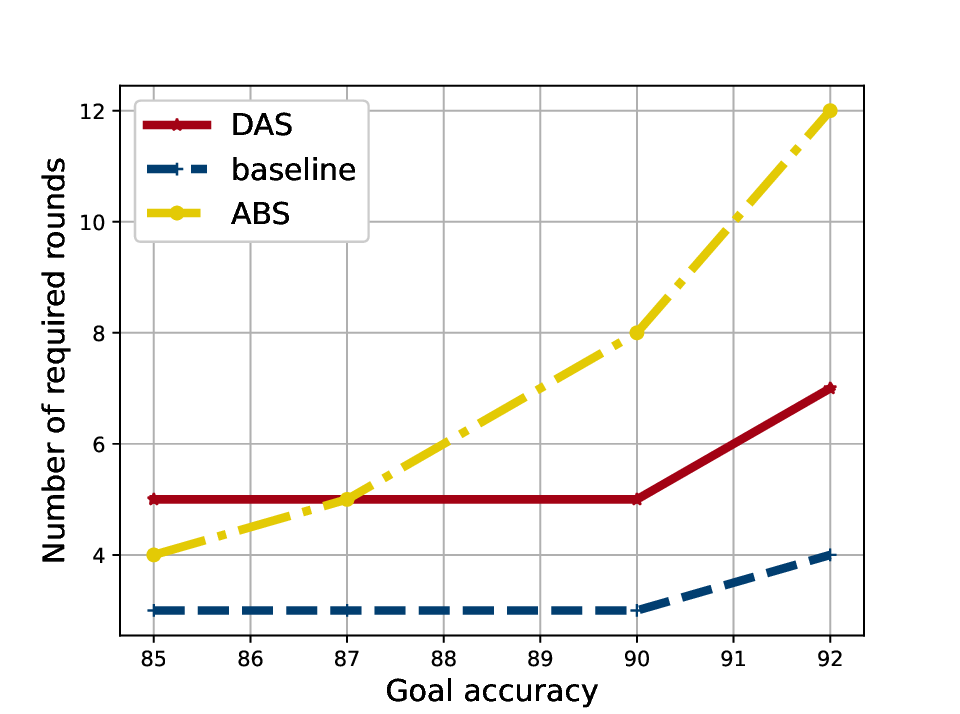}
		\caption{$s=100k$}
		\label{s100}
	\end{subfigure} \hspace{3mm}
	\begin{subfigure}[t]{0.3\textwidth}
		\includegraphics[width=1.4\linewidth]{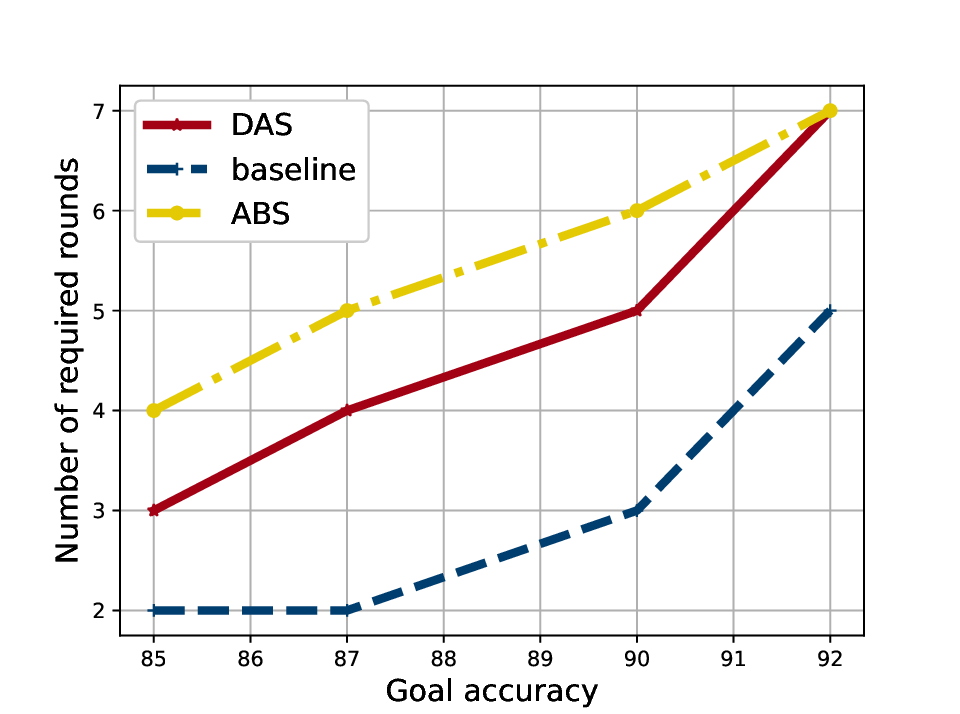}
		\caption{$s=150k$}
		\label{s150}
	\end{subfigure} \hspace{3mm}
	\begin{subfigure}[t]{0.3\textwidth}	
		\includegraphics[width = 1.4\linewidth]{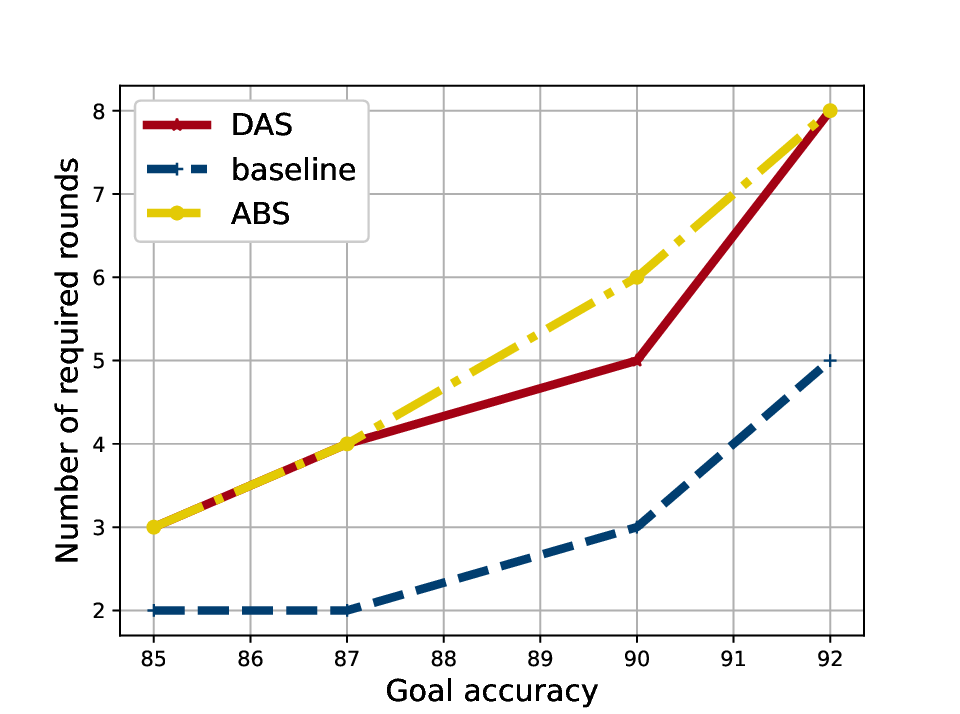}
		\caption{$s=200k$}
		\label{s200}
	\end{subfigure}
	\caption{The impact of the model size on the required number of rounds to achieve the desired accuracy using the CNN model.}
	\label{cnnsize}
	%\end{center}
\end{figure}

\begin{figure}[h]
	%\centering
	%\begin{center}
	%\hspace{0.07 in}
	\begin{subfigure}[t]{0.3\textwidth}
		\includegraphics[width=1.4\linewidth]{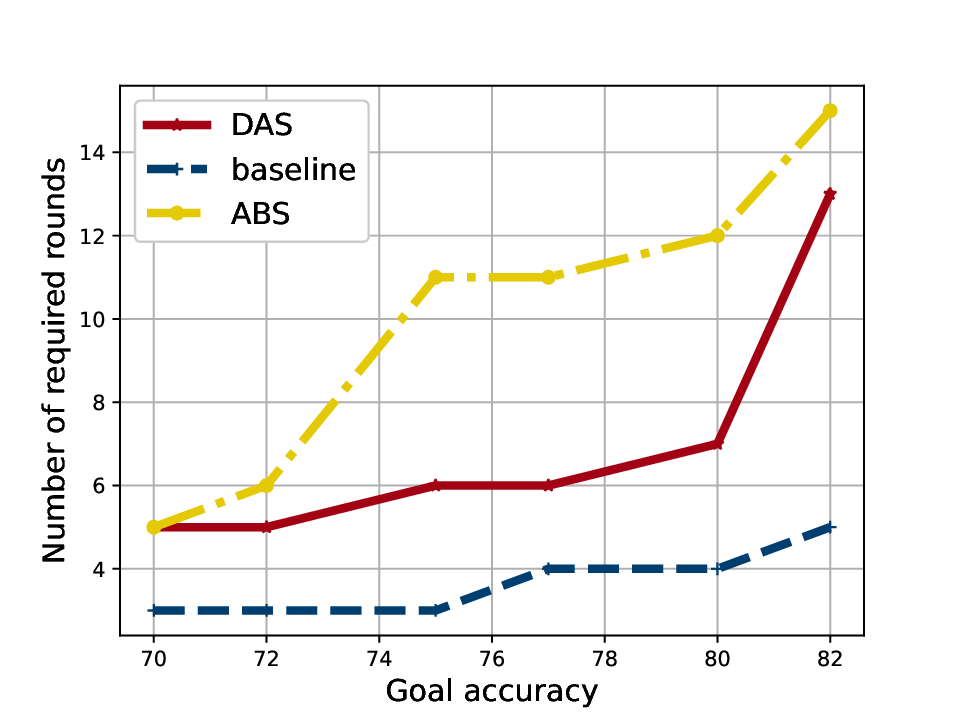}
		\caption{$s=100k$}
		\label{mlps100}
	\end{subfigure} \hspace{3mm}
	\begin{subfigure}[t]{0.3\textwidth}
		\includegraphics[width=1.4\linewidth]{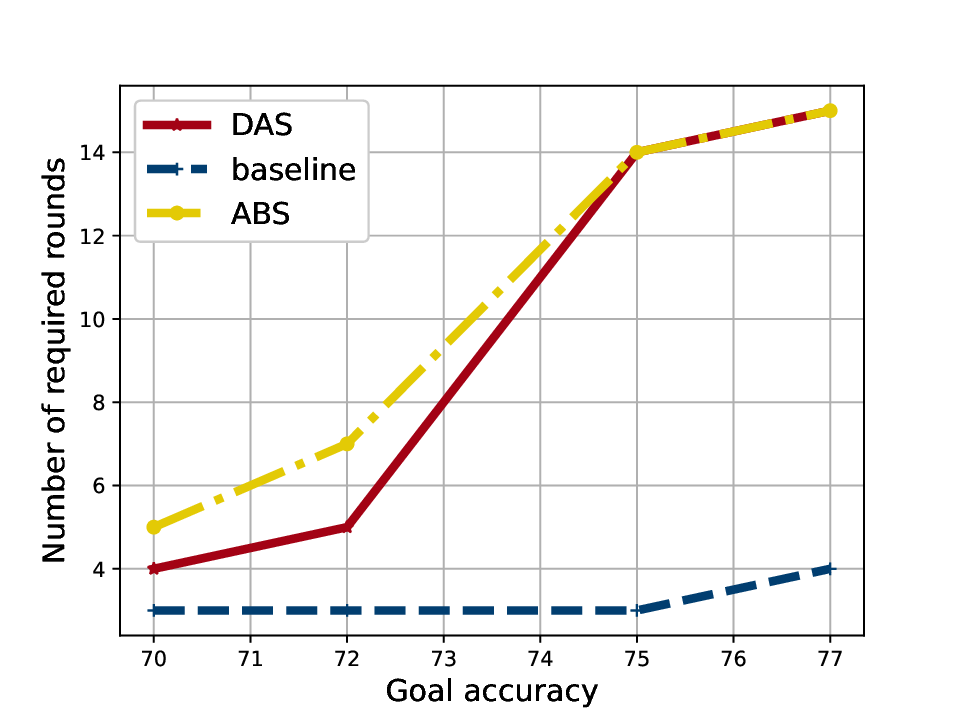}
		\caption{$s=150k$}
		\label{mlps150}
	\end{subfigure} \hspace{3mm}
	\begin{subfigure}[t]{0.3\textwidth}	
		\includegraphics[width = 1.4\linewidth]{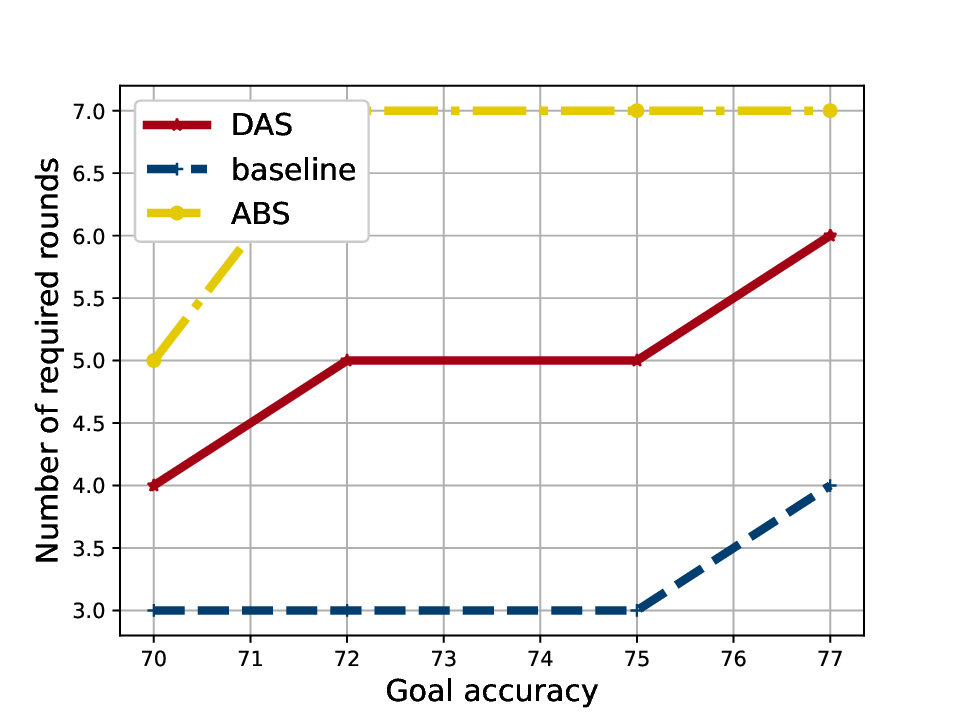}
		\caption{$s=200k$}
		\label{mlps200}
	\end{subfigure}
	\caption{The impact of the model size on the required number of rounds to achieve the desired accuracy using the MLP model.}
	\label{mlpsize}
	%\end{center}
\end{figure}

\begin{figure}[htb]

	\begin{subfigure}[t]{0.3\textwidth}
		\includegraphics[width=1.4\linewidth]{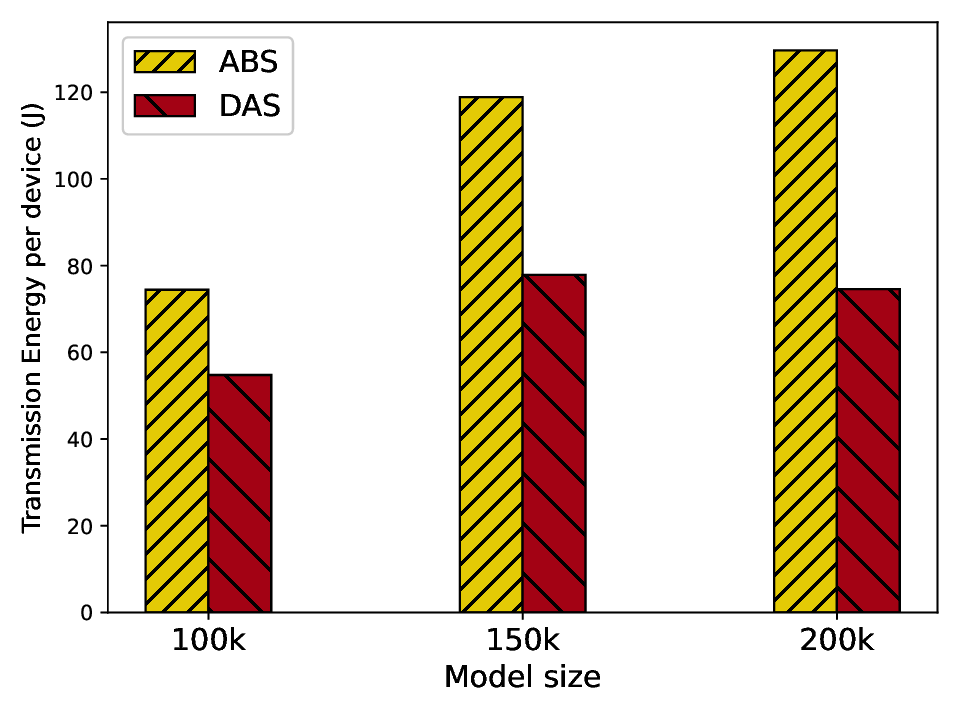}
		\caption{Energy per device}
		\label{mlp_energy}
	\end{subfigure} \hspace{18mm}
	\begin{subfigure}[t]{0.3\textwidth}
		\includegraphics[width=1.4\linewidth]{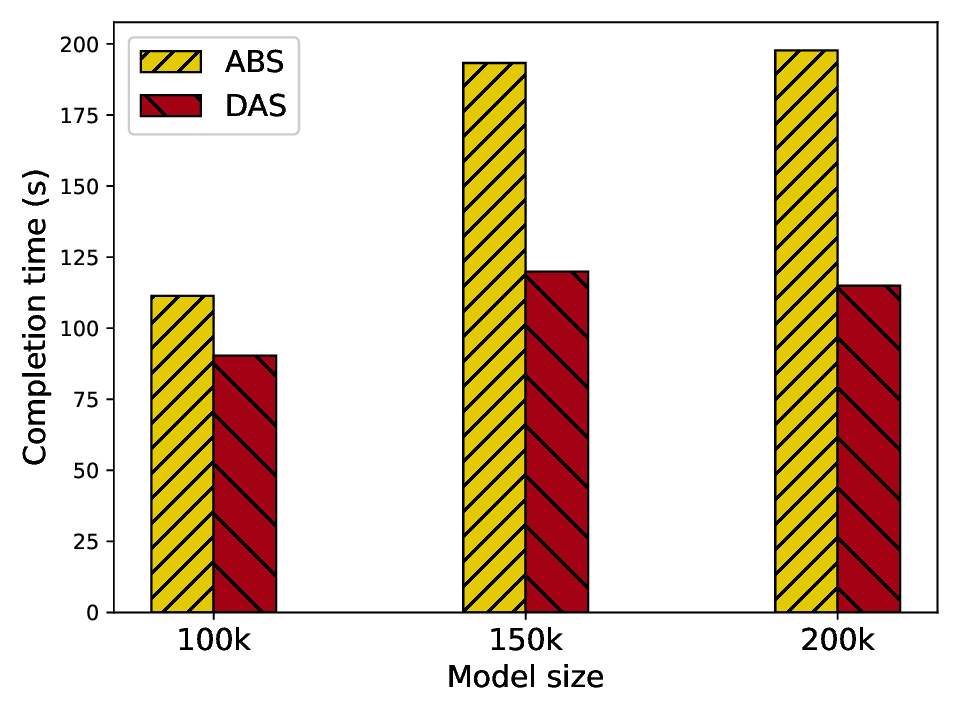}
		\caption{Completion time}
		\label{mlp_time}
	\end{subfigure} \hspace{3mm}

	\caption{Energy per device and completion time for training the CNN model  for a goal accuracy of 92\%}
	\label{cnn_time_energy}
\end{figure}

\begin{figure}[htb]

	\begin{subfigure}[t]{0.3\textwidth}
		\includegraphics[width=1.4\linewidth]{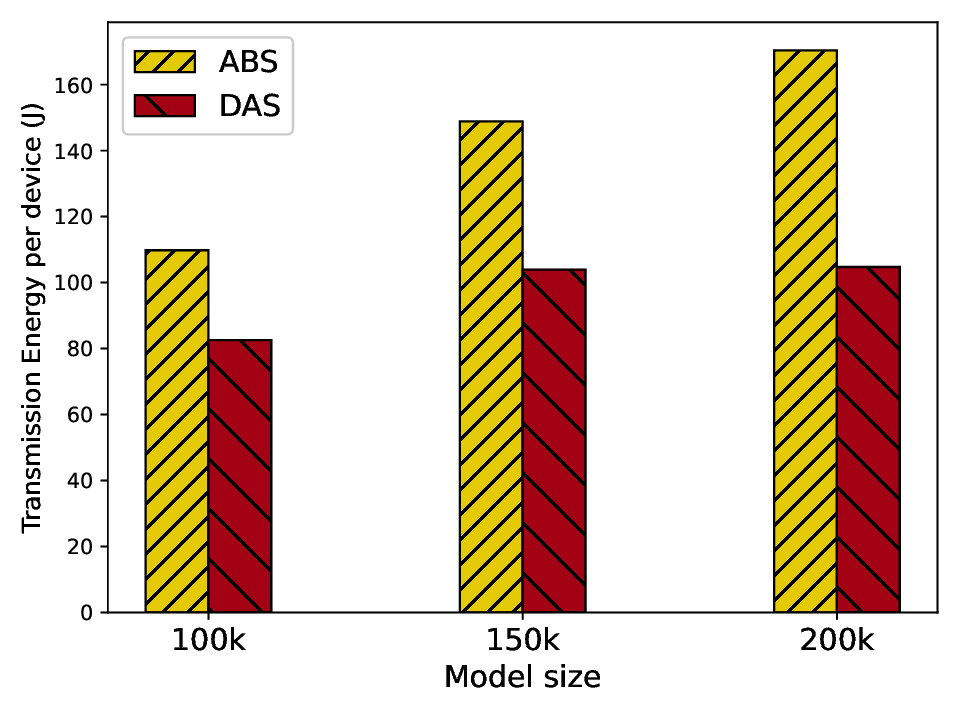}
		\caption{Energy per device}
		\label{mlp_energy1}
	\end{subfigure} \hspace{18mm}
	\begin{subfigure}[t]{0.3\textwidth}
		\includegraphics[width=1.4\linewidth]{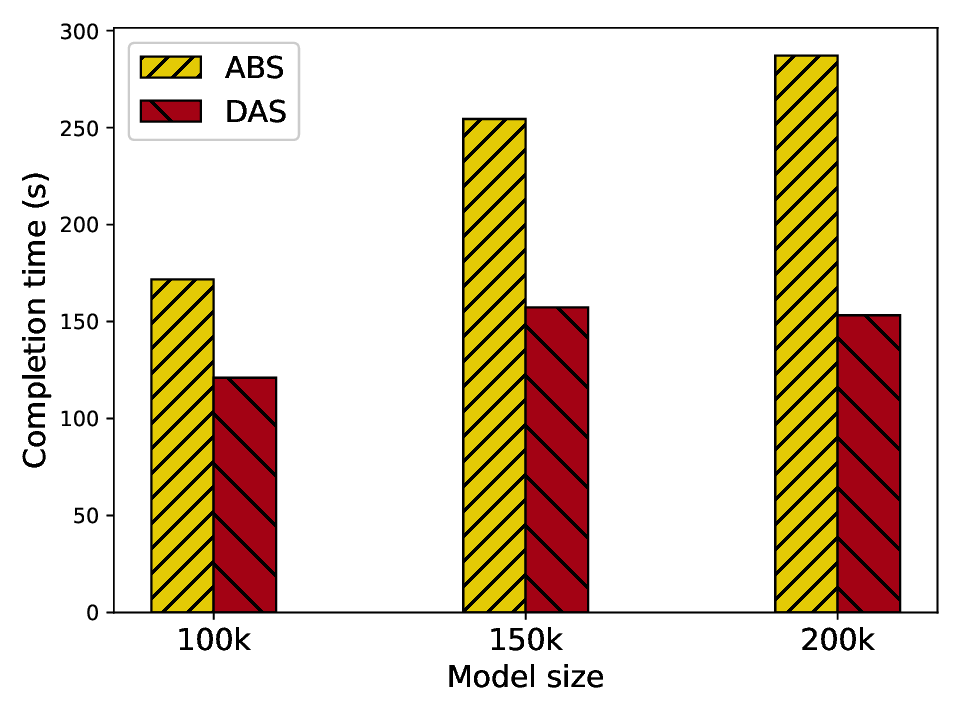}
		\caption{Completion time}
		\label{mlp_time1}
	\end{subfigure} 

	\caption{Energy per device and completion time for training the MLP model for a goal accuracy of 77\% }
	\label{mlp_time_energy}
\end{figure}

\begin{figure}[t]
	%\centering
	%\begin{center}
	%\hspace{0.07 in}
	\begin{subfigure}[t]{0.3\textwidth}
		\includegraphics[width=1.4\linewidth]{figures/cnnsize100.eps}
		\caption{$E=1$}
		\label{e1cnn}
	\end{subfigure} \hspace{3mm}
	\begin{subfigure}[t]{0.3\textwidth}
		\includegraphics[width=1.4\linewidth]{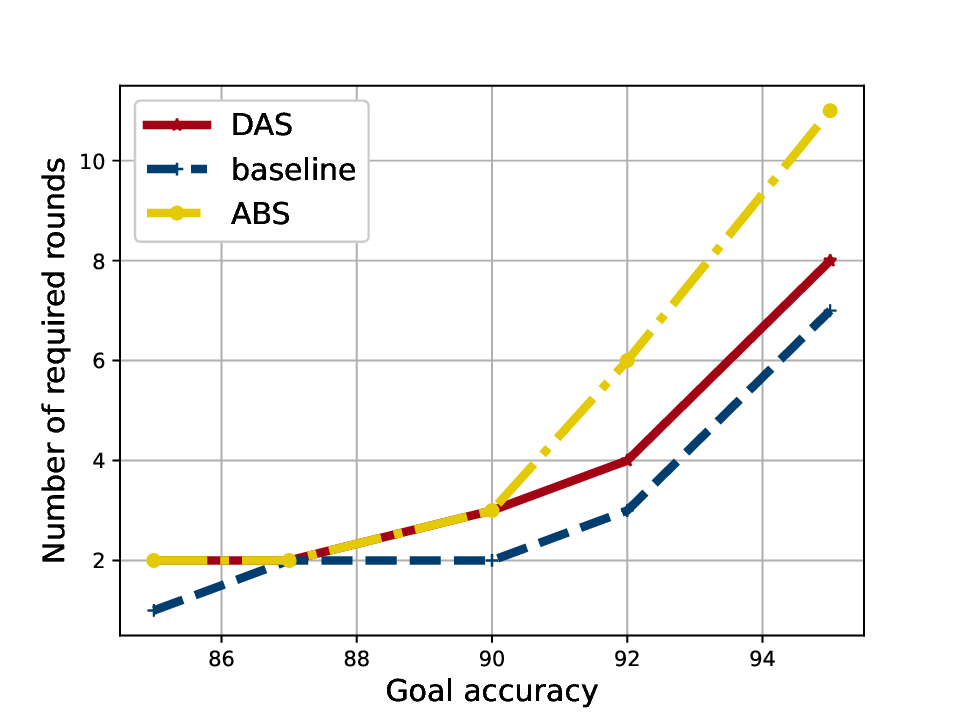}
		\caption{$E=2$}
		\label{e2cnn}
	\end{subfigure} \hspace{3mm}
	\begin{subfigure}[t]{0.3\textwidth}	
		\includegraphics[width = 1.4\linewidth]{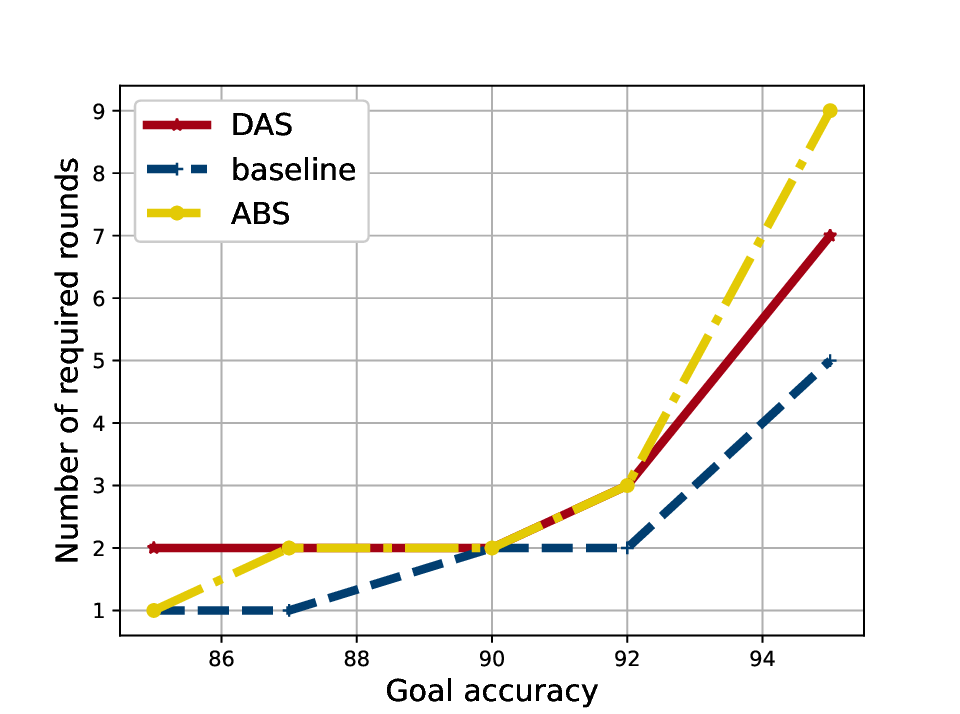}
		\caption{$E=3$}
		\label{e3cnn}
	\end{subfigure}
	\caption{The impact of the number of local epochs on the required number of rounds to achieve the desired accuracy using the CNN model.}
	\label{cnnepochs}
	%\end{center}
\end{figure}

\begin{figure}[htb]

	\begin{subfigure}[t]{0.3\textwidth}
		\includegraphics[width=1.4\linewidth]{figures/mlpsize100.eps}
		\caption{$E=1$}
		\label{e1mlp}
	\end{subfigure} \hspace{3mm}
	\begin{subfigure}[t]{0.3\textwidth}
		\includegraphics[width=1.4\linewidth]{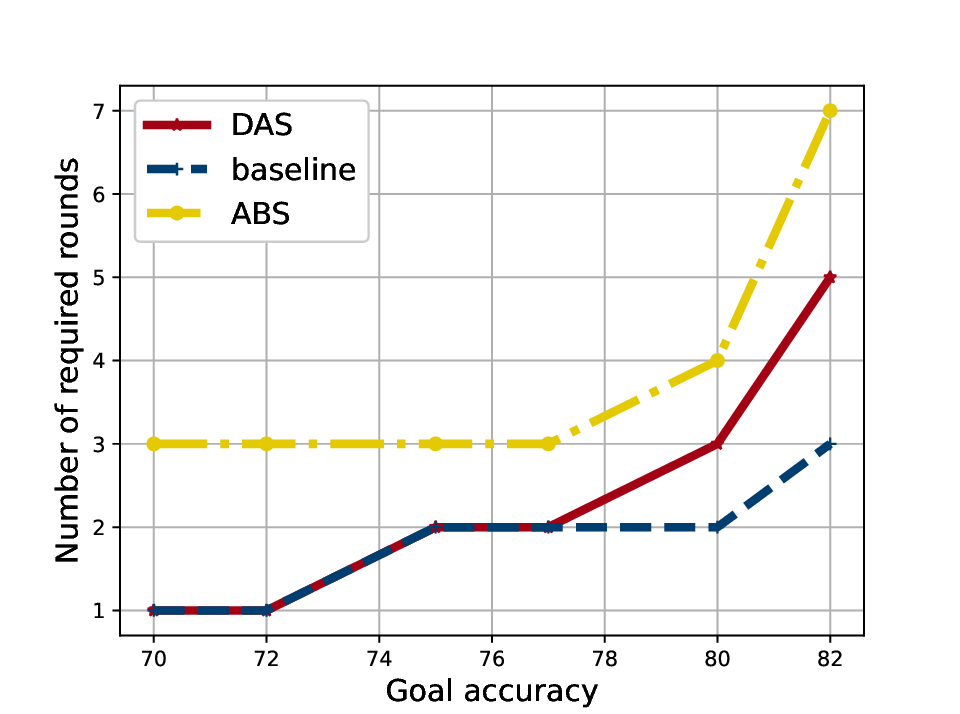}
		\caption{$E=2$}
		\label{e2mlp}
	\end{subfigure} \hspace{3mm}
	\begin{subfigure}[t]{0.3\textwidth}	
		\includegraphics[width = 1.4\linewidth]{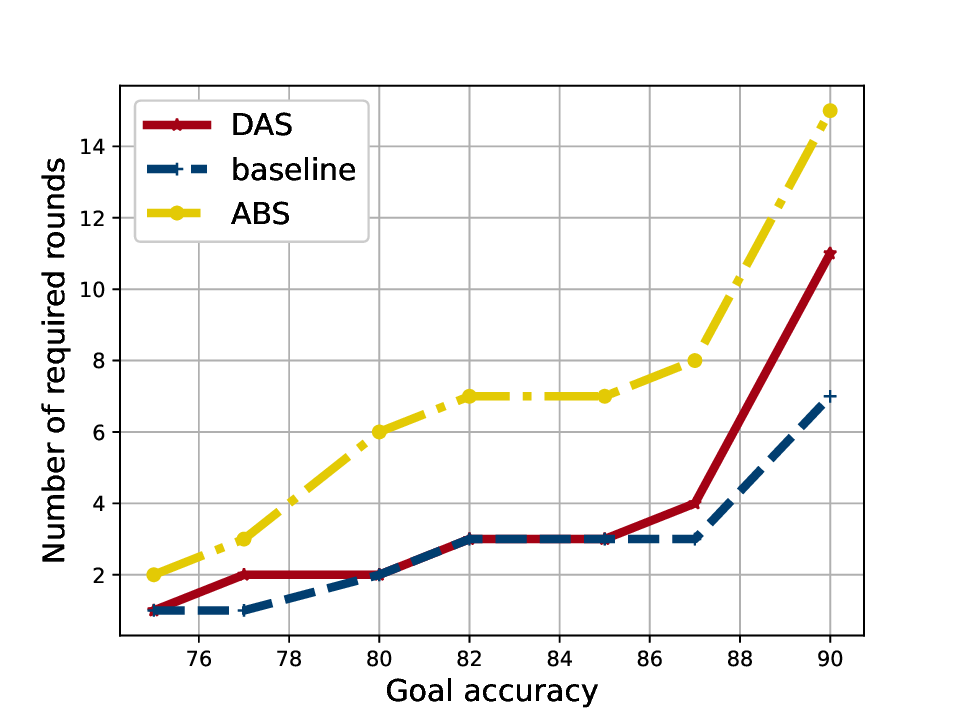}
		\caption{$E=3$}
		\label{e3mlp}
	\end{subfigure}
	\caption{The impact of the number of local epochs on the required number of rounds to achieve the desired accuracy using the MLP model.}
	\label{mlpepochs}
\end{figure}

\begin{figure}[htb]

	\begin{subfigure}[t]{0.3\textwidth}
		\includegraphics[width=1.4\linewidth]{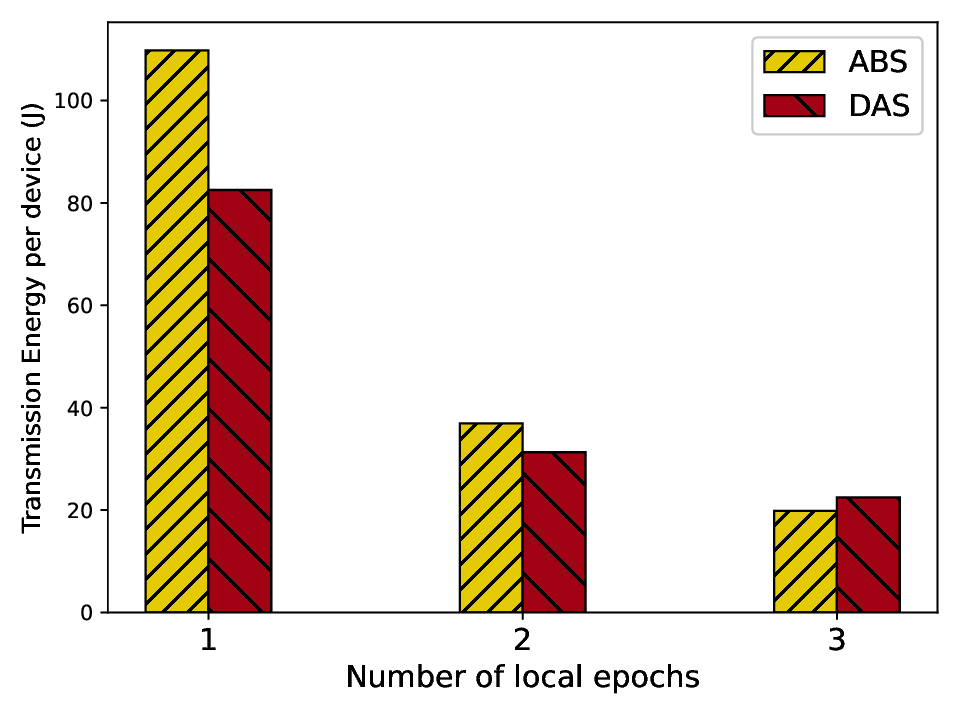}
		\caption{Energy per device}
		\label{cnn_energy2}
	\end{subfigure} \hspace{18mm}
	\begin{subfigure}[t]{0.3\textwidth}
		\includegraphics[width=1.4\linewidth]{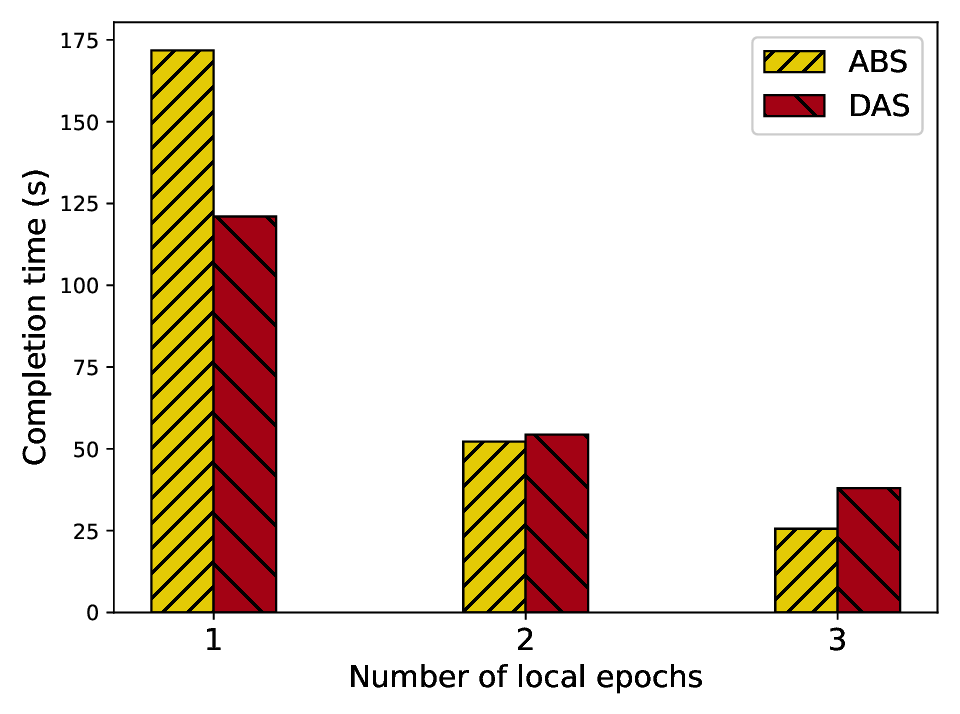}
		\caption{Completion time}
		\label{cnn_time2}
	\end{subfigure}

	\caption{Energy per device and completion time for training the CNN model  for a goal accuracy of 92\%}
	\label{cnn_time_energy_epochs}
\end{figure}

\begin{figure}[htb]

	\begin{subfigure}[t]{0.3\textwidth}
		\includegraphics[width=1.4\linewidth]{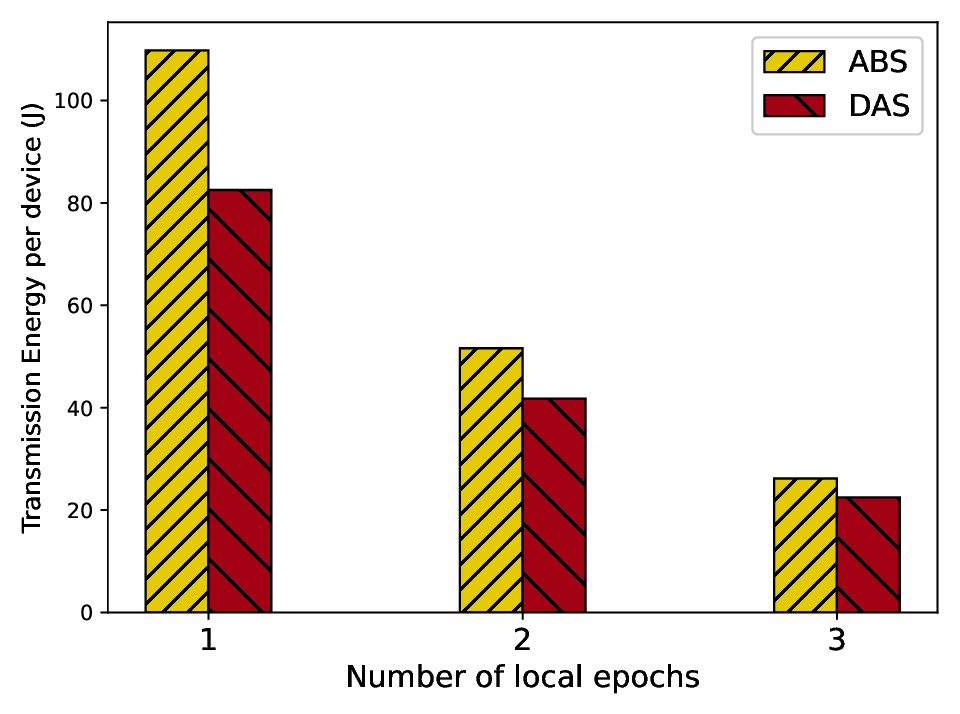}
		\caption{Energy per device}
		\label{mlp_energy2}
	\end{subfigure} \hspace{18mm}
	\begin{subfigure}[t]{0.3\textwidth}
		\includegraphics[width=1.4\linewidth]{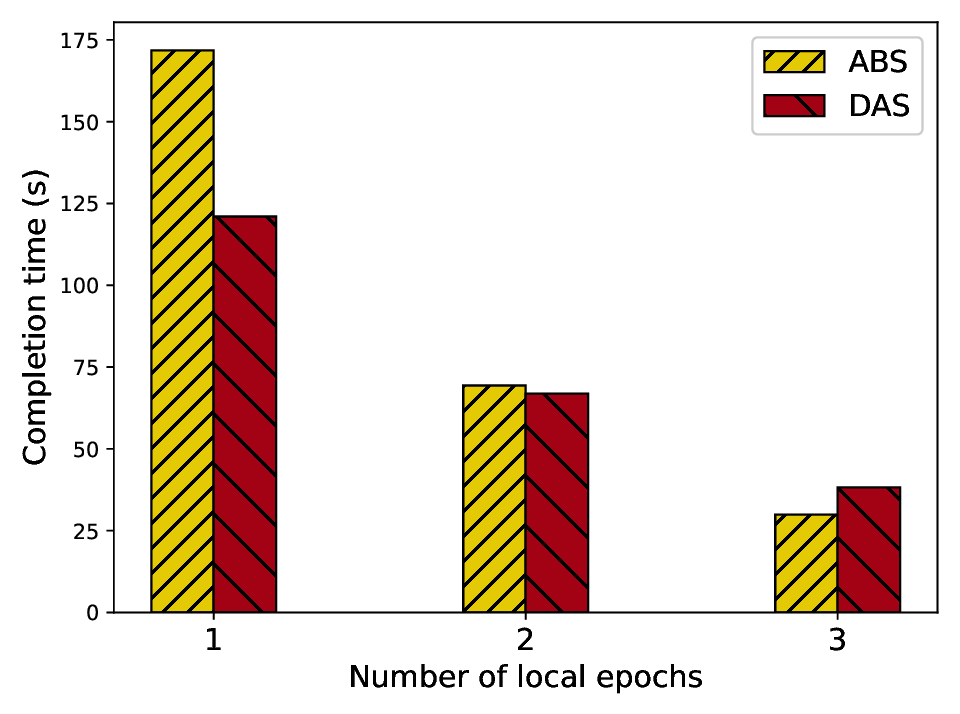}
		\caption{Completion time}
		\label{mlp_time2}
	\end{subfigure} 

	\caption{Energy per device and completion time for training the MLP model for a goal accuracy of 77\% }
	\label{mlp_time_energy_epochs}
\end{figure}
\subsection{Diversity Index Evaluation}\label{subsec:index1}
To test the proposed index, we train both models (i.e., CNN and MLP), using  the  FedAvg  algorithm~\cite{f3} over a  total of 15 rounds. {\revision Since MNIST is essentially an image classification task, and we have generated highly unbalanced datasets where several devices only have a subset of the target classes, we used Gini-Simpson index to evaluate the datasets diversity.} We set the weights of the index equally  to $1/3$.  We compare DAS performance to random selection-based scheduling.

To clearly illustrate the efficiency of the diversity index in devices' selection, we stress-test the selection by limiting the number of selected devices. Furthermore, the trade-off  between  local  update  and global aggregation imposes the evaluation of the performance when varying the number of local iterations.

As a first experiment, we limited the number of selected devices to 3, 5 and 7. Fig. \ref{cnn_mlp_clients} shows the obtained accuracy throughout training rounds. The average accuracy obtained using DAS is significantly higher across different simulations. Thus, it is clear that using the diversity index as a criteria for scheduling devices allows to accelerate the learning significantly, especially when the number of devices that can be scheduled is low. 

In a second experiment, in Fig. \ref{cnn_mlp_epochs}, we fix the number of selected devices to 7, and we vary the number of local epochs $E \in \{ 1, 2, 3 \}$ while choosing a random scheduling algorithm or DAS algorithm. Fig. \ref{cnn_mlp_epochs} illustrates the obtained results for the CNN and the MLP models. Adding more local computations allows significant gains in communication and in accuracy, especially for the MLP model. We notice that the data-aware device selection still surpasses random selection in these simulations when adding more local computation.
All in all, these results validate our hypothesis involving the importance of dataset diversity.

\subsection{Effect of Model Size}\label{subsec:size1}
%\textit{1) Effect of model size:}
%\newline

Fig. \ref{cnnsize} and Fig.\ref{mlpsize} show that using DAS algorithm, the number of required communication rounds to reach the desired accuracy is always lower than (or at least equal to)  the one needed using ABS. For both models, the size can affect the convergence speed of the learning. 
It should be noted that ABS tends to select more devices in early communication rounds, which can yield higher accuracy than DAS similarly to what is shown in Fig.\ref{s100}. Nonetheless, ABS gives higher priority to devices that did not participate, which leads to a decrease in the number of devices that can be selected, and thus a slower convergence over the following rounds.  

Indeed, the baseline scenario requires less rounds to reach high accuracy levels, nonetheless, it is hard to scale for large number of devices in rapidly changing environments. In fact, while Fig. \ref{mlp_time_energy} and Fig.\ref{cnn_time_energy} show that the ABS and DAS are comparable in terms of energy and completion time, the consumed energy and time only represent a fraction of those of the baseline scenario requirements. This is mainly due to scheduling only a fraction of the devices, which for both algorithms did not exceed 20\%.

Considering a goal accuracy of 77\% for the MLP model and 92\% for the CNN, the values in Fig.\ref{cnn_time_energy} and Fig.\ref{mlp_time_energy}, represent gains in energy compared to the baseline as follows: 
On total, the consumed energy per device for the ABS represents a gain of 68,85\% on average for training the MLP model and 76,56\% for the CNN model.  Even higher gains are achieved for DAS, where a gain of 78,86\% is reached when training the MLP model and 84,96\% CNN.
In terms of completion time, the required time across different experiments for DAS is significantly less than the required time for ABS. These results are consistent with the number of required rounds for training presented in Fig. \ref{cnnsize} and Fig.\ref{mlpsize}. Additionally, the ABS algorithm selects more devices in the first few rounds, and much less devices in the later rounds, thus leading to longer rounds in the beginning of the training, and shorter rounds later. 

These results show that the careful selection of participating devices jointly with the optimized  bandwidth allocation, {\revision make DAS scalable in terms of time and energy when training large models.} 

\subsection{Local Computation}\label{subsec:comp1}
%\textit{2) Local computation:}
%\newline

In this section, we study the effect of increasing the computation per device. We fix $s = 100 $kbits, and add more local computation per client on each round. 

By increasing the number of local epochs $E$, we take full advantage of available parallelism on the client hardware, which leads to higher accuracy on the test set with less communication rounds. However, in previous work \cite{caldas_leaf_2019, f3}, long local computation may lead to the divergence training loss. Furthermore, due to the changing environment due to the mobility of the devices, it is hard to plan ahead the communication rounds for large $E$.  Therefore, we limit the experiments to $(1, 2 ,3)$ local epochs.

Fig. \ref{cnnepochs} and Fig. \ref{mlpepochs} show that adding more local epochs per round can produce a dramatic decrease in communication costs. We see that increasing the number of local epochs $E$ can benefit largely from DAS, as it achieves a closer behaviour to the baseline, especially for the less powerful MLP model. 

Fig. \ref{cnn_time_energy_epochs} and Fig. \ref{mlp_time_energy_epochs} show how trading frequent communication with more local computation has several benefits, as it reduces the required transmission energy, as well as the FEEL completion time. 
Similarly to the previous experiments, we considered a goal accuracy of 77\% for the MLP model and 92\% for the CNN, the gains in energy compared to the baseline when increasing the number of local epochs are as follows: 
On total, the consumed energy per device for the ABS represents gains of 83.5\% and 95.97\%  for the MLP when training for 2 epochs and 3 epochs respectively 88.19\%, 96,95\% on average for training the CNN model. Even higher gains are achieved for DAS, where a gain of 86.65\% and 96.54\% is reached when training the MLP model and 90\% and 96,54\% for the CNN model. These results are consistent with the fraction of selected devices which does not exceed 20\% on average.

We noticed that when increasing the number of local epochs, the required completion time using DAS becomes slightly higher than the required completion time for ABS. This is mainly due to prioritizing devices that have larger datasets, which leads to longer training duration. Nevertheless, the difference can be seen as marginal when considering the gain in energy.

%These results show that the careful selection of participating devices and the resource allocation, make the DAS scalable. 

\section{Conclusion}
\label{sec:conclusion}
In this paper, we have investigated the problem of devices scheduling in federated edge learning by formulating the following question: Can the use of a suitable diversity index help achieve a better accuracy in fewer rounds? 
To answer this question, we consider data properties as the key motor of the selection of devices, as we designed a diversity index which can be adapted to a wide variety of use-cases. Additionally, we integrated the diversity index in a novel scheduling strategy in wireless networks, where the completion time and energy efficiency of the transmission  are also of high importance. To this end, we derived the time and energy consumption models for FEEL based on devices and channels properties. With these models, we have formulated a joint  selection and bandwidth allocation problem, aiming to minimize a multi-objective function of the completion time and the total transmission energy, while balancing with a goal to maximizing the diversity of the selected devices. 
We have proposed to solve this problem through an iterative algorithm that starts with the selection of the devices and then allocates the bandwidth. Through extensive evaluations, we proved the importance of the data properties in FEEL and the efficacy of the diversity index.
Furthermore, we showed that our proposed scheduling algorithm can  effectively reduce the number of required rounds to achieve high accuracy levels especially for large models, which results also in savings in time and energy.

\section*{ACKNOWLEDGEMENT}
\label{sec:acknowledgement}
The authors would like to thank the Natural Sciences and Engineering Research Council of Canada, for the financial support of this research.

%================== references  =================

\bibliography{./includes/references}

\end{document}